\documentclass[11pt]{amsart}
\usepackage{hyperref}
\usepackage[alphabetic]{amsrefs}
\usepackage{geometry}                
\usepackage{graphicx}
\usepackage{amssymb}
\usepackage{epstopdf}

\newcommand{\CZ}{\mathcal{Z}}
\newcommand{\CC}{\mathcal{C}}

\title{Resurgence and Holomorphy:\\ From Weak to Strong Coupling}

\author{Aleksey Cherman}
\address{Fine Theoretical Physics Institute, Department of Physics, University of Minnesota, Minneapolis, MN 55455, USA}
\email{acherman@umn.edu}

\author{Peter Koroteev} 
\address{Perimeter Institute for Theoretical Physics, Waterloo, ON N2L2Y5, Canada}
\email{pkoroteev@perimeterinstitute.ca}

\author{Mithat \"Unsal}
\address{Department of Physics, North Carolina State University, Raleigh, NC 27695 USA} 
\email{unsal.mithat@gmail.com}

\date{\today}                                     

\begin{document}

\begin{abstract}
We analyze the resurgence properties of 
finite-dimensional exponential integrals which are  prototypes for partition functions in quantum field theories.  In these simple examples, we demonstrate 
that perturbation theory, even at arbitrarily weak coupling, fails as the argument of the coupling constant is varied. It is  well-known that perturbation theory also fails at stronger coupling. We show that these two failures are actually intimately related. The formalism of resurgent transseries, which takes into account global analytic continuation properties, fixes both problems, and  provides an arbitrarily accurate description of exact result for any value of coupling.  This means that  
 strong coupling results can be deduced by using merely weak coupling data. 
Finally, we give another perspective on resurgence theory by showing that the monodromy properties of the weak coupling results are in precise agreement with the monodromy properties of the strong-coupling expansions,  obtained using analysis of the holomorphy structure of Picard-Fuchs equations. 
\end{abstract}

\begin{flushright}
FTPI-MINN-14/30,  UMN-TH-3403/14
\end{flushright}

\bigskip

\maketitle
\tableofcontents

\section{Introduction}
In quantum mechanics (QM) and in quantum field theory (QFT)  weak coupling  perturbative expansions
and semi-classical expansions are often the only available computational techniques.   Yet perturbative expansions almost always yield asymptotic series, with a zero radius of convergence.  While these asymptotic expansions are often very useful when the coupling is small, $g \ll 1$, we also often want to understand what happens when $g \gg 1$ or $g \sim 1$.  Can perturbative expansions say anything useful once $g \gtrsim 1$?  Even if we are content to focus on the $g \ll 1$ regime, we may want to understand non-perturbative effects which contribute as e.g. $e^{-A/g}, A \in \mathbb{R}^{+}$, which might naively appear to be entirely invisible in power series expansions in $g$.  Do perturbative expansions somehow contain quantitative information about these effects? Recently there has been a revival of interest in these classic issues, driven by developments in the application of \`Ecalle's resurgence theory \ycite{Ecalle:1981}\footnote{See also the early work by Dingle \cite{Dingle_asymptotics}.} to 0d matrix models, quantum mechanics, quantum field theory, and string theory, see e.g.~\cite{Garoufalidis:2010ya,Aniceto:2011nu,Marino:2012zq,Argyres:2012ka,Argyres:2012vv,Dunne:2012ae,Dunne:2012zk,Aniceto:2013fka}, following the earlier work in e.g. \cite{berry1990hyperasymptotics,berry1991hyperasymptotics,delabaere1997unfolding,ZinnJustin:2004cg,ZinnJustin:2004ib}.   For a recent introduction to resurgence theory from a mathematical perspective see e.g. \cite{sauzin2014introduction}.

In many examples, resurgence theory has been found to give a deep understanding of the structure of divergences in perturbative expansions, and has illuminated surprising quantitative relationships between perturbative and non-perturbative effects in weakly-coupled theories.

Here we explore some implications of resurgence theory in the context of  ordinary exponential integrals, which can be considered to be zero dimensional prototypes for Euclidean path integrals and partition functions.   The property that the perturbative expansions for these toy models are divergent asymptotic series  is also shared with  path integrals   defining QM and QFT systems, albeit in a simplified form.  
The usual perturbative expansion around the perturbative saddle can fail to be close to the exact result in two different ways: 
\begin{itemize} 
\item {Perturbation theory (around the perturbative saddle)  fails to accurately approximate the exact result at strong coupling where $g \gg 1$. 
}  
\item {Even when the magnitude of  the coupling  $|g|$  is {\it arbitrarily small}, 
perturbation theory (around the perturbative saddle)  fails to accurately approximate the exact result at  generic $\theta= \arg g$. } 
\end{itemize}
The first of these issues is of course obvious, while the second is not as widely appreciated.   The surprising fact emphasized in this paper is that these two issues are actually deeply related, and the root cause for these two failures of naive perturbative expansions is identical.  The existence of non-perturbative saddles in the problem is not taken into account in naive perturbation theory.   
 Consistently incorporating \emph{all} of the saddle points in the system, i.e, building a resurgent transseries, 
with the correct analytic continuation properties in the coupling, we find that:  
\begin{itemize} 
\item{
At arbitrarily small $|g| \ll 1$, 
resurgent transseries   approximate the exact result at  generic $\theta= \arg g$ to an  arbitrary accuracy,  for the whole range of  $\theta= \arg g$.} 
\item{  The same  weak-coupling  resurgent transseries data can be used  to make arbitrarily  accurate strong-coupling predictions.  Hence the solution of the weak coupling and strong couplings problems is one and the same, as argued above.}
\end{itemize}

These simple observations provide a new perspective (compared to, e.g. \cite{Witten:2010cx, Witten:2010zr}) on analytic continuation in ordinary and path integrals, and gives us hope that certain strong coupling issues can in fact be addressed and understood at weak coupling!   The rationale is as follows:
  Working at weak coupling, one can turn exponentially small  (recessive) contributions $e^{-A/g}$ into exponentially large (dominant) contributions $e^{+A/|g|}$ using analytic continuation in  $\arg g$, bringing these non-perturbative effects into a much more visible form. 
On the other hand, when  $\arg g=0$ and we  dial $g$ from weak coupling to strong coupling, 
 the  $e^{-A/g}$ factors in the transseries also become order one, instead of being exponentially suppressed with respect to the perturbative contribution.  The combination of the perturbative and non-perturbative terms in the transseries gives the strong coupling result to arbitrary accuracy.  It is crucial to  note that this idea is different from knowing perturbative expansions on the strong and weak coupling sides, and finding interpolating functions connecting the two expansions, which has recently been discussed in a series of works \cite{Sen:2013oza,Beem:2013hha,Pius:2013tla,Honda:2014bza,Alday:2013bha,Banks:2013nga}.  It is also conceptually different from techniques like variational perturbation theory, where one finds a scheme to extrapolate the perturbative series around a zero-action saddle point toward strong coupling, described in e.g. \cite{Feynman:1986ey,kleinert2010converting}.   The only data that we use 
 to predict the observables at arbitrary coupling is the weak coupling transseries.  
 Thus, resurgence theory allows us 
 to learn about the strong coupling effects from weak coupling information, with the global analytic continuation properties of the problem treated in a systematic way.  

The reason that the weak coupling resurgent expansion is capable of  reproducing the exact result at arbitrary coupling is that resurgent transseries have the same monodromy properties
under global analytic continuation as exact solutions, by construction.  This is to be contrasted with naive perturbation theory, which manifestly does not have the correct monodromy properties.  Remarkably, the monodromy properties of the problem can also be derived from an examination of the strong coupling expansion,  which gives \emph{convergent} series.  This can be done without appealing to the exact solutions.  Instead, using a holomorphy analysis of the structure of Picard-Fuchs equations  
(which can be viewed as analogs of Ward identities in QFT), we derive the monodromy matrices characterizing the analytic continuation properties of our toy examples.  We show that these strong-coupling monodromy matrices contain precisely the same information as the weak-coupling monodromy matrices. This provides us with a useful complementary perspective on resurgence theory.  

We hope that the techniques we highlight here in simple toy models may be of use in quantum mechanics and QFTs, at least in cases where there are no phase transitions as one moves from weak coupling to strong coupling.

\section{Periodic Potential and two failures of perturbation theory}
\begin{figure*}[th]
  \centering
\includegraphics[width=\textwidth]{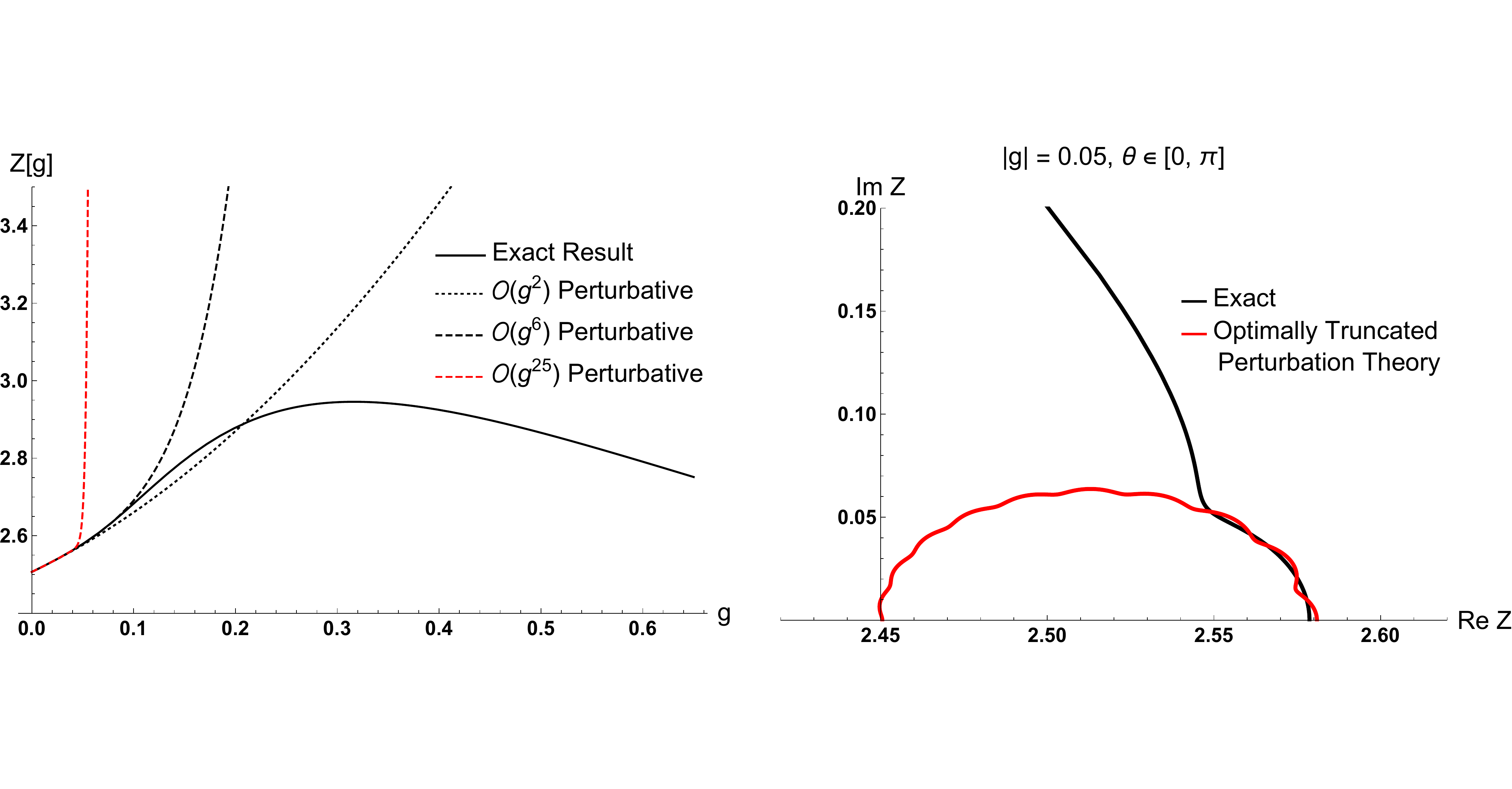}
  \caption{(Color Online.) The left-hand plot illustrates the (unsurprising) failure of naive perturbation theory to accurately approximate the exact `partition function' $\mathcal{Z}(g)$ once $g$ leaves the $g \ll 1$ domain.  The right-hand plot illustrates the less familiar point that even when $|g| \ll 1$, naive perturbation theory fails to give correct results once $\theta \equiv \arg g \neq 0$.  }
  \label{fig:NaivePerturbationTheory}
\end{figure*}

Consider the following exponential  integral
\begin{equation}
\mathcal{Z}(g) = \frac{1}{\sqrt{g}} \int\limits_{-\pi/2}^{\pi/2} dx \, e^{-\frac{1}{2g} \sin(x)^2} 
=  \int\limits_{-\pi/(2\sqrt{g})}^{+\pi/(2\sqrt{g})} dy\, e^{-\frac{1}{2g} \sin(y g^{1/2} )^2}\,,
\label{eq:0dPerodicPotPartFun}
\end{equation}
which formally can be thought of as a partition function of a self-interacting $0+0$ dimensional QFT\footnote{In cases where expressions like \eqref{eq:0dPerodicPotPartFun} appear via e.g. dimensional reduction from an actual QFT, the quantity we are calling $g$ is often the square of the coupling constant of the QFT.}. 
In the above formula the second expression is canonically normalized such that the Gaussian terms are $g$-independent.  In this particular example, $\mathcal{Z}(g)$ can be explicitly evaluated in terms of known functions:
\begin{equation}
\mathcal{Z}(g) = \frac{\pi}{\sqrt{g}} e^{-\frac{1}{4g}}I_0\left(\frac{1}{4g}\right)\,.
\label{eq:SolParFuncReal}
\end{equation}
However, in the rest of the section we shall generally proceed assuming that we do not know the exact answer, and \eqref{eq:SolParFuncReal} will be used as a consistency check for our results.

To motivate the subsequent discussion, we note that when $|g| \ll 1$, it is tempting to think that the standard perturbative power series approximation  (around the perturbative saddle)
\begin{align}
\mathcal{Z}(g) \overset{?}{\simeq} \sqrt{2\pi} \left[1+ \frac{g}{2} + \frac{9 g^2}{8} + \frac{75 g^3}{16} + \frac{3675 g^4}{128}+\frac{59535 g^5}{256} +\frac{2401245 g^6}{1024}+ \cdots\right]
\label{eq:purePT}
\end{align}
should accurately approximate \eqref{eq:0dPerodicPotPartFun}.  (The derivation of this expression is given in the following section.)  Indeed, if the number of terms in \eqref{eq:purePT}  is fixed to $n_*$, then for any value of $\arg g$, one can see that if $|g|$ is sufficiently small, the $\mathcal{O}(g^{n+1})$ term is parametrically smaller in absolute value than the $\mathcal{O}(g^{n})$ term for any $0\le n \le n_*$.  Hence, so long as $|g| \ll 1$, the standard criterion for the reliability of a perturbative expansion is satisfied, and one might thus expect that \eqref{eq:purePT} will accurately approximate \eqref{eq:0dPerodicPotPartFun} for any $\arg g$.

So long as $\arg g = 0$, this standard intuition is correct.  However, as illustrated\footnote{The exploration of the behavior of asymptotics as a function of $\arg g$  dates all the way to Stokes \cite{Stokes1864}, and was popularized in recent years starting with \cite{berry1991asymptotics}.}  by the left plot in Fig.~\ref{fig:NaivePerturbationTheory}, \eqref{eq:purePT} fails badly once $\arg g$  deviates from zero!  Fig.~\ref{fig:NaivePerturbationTheory} also verifies the less surprising fact that the naive perturbative approximation represented by \eqref{eq:purePT} does not work well once $g \sim 1$.  In the  subsequent sections, we explain the reasons for these failures of perturbation theory from several perspectives, and explain that actually the two failures are interrelated. 

\subsection{Weak Coupling Approach}\label{Sec:TrigWeakCoupling}
Most of the results in this section are well-known and have previously appeared in the literature in e.g.~\cite{Dunne:2012ae,Cherman:2014ofa}, but we review them here to keep our discussion self-contained.  

The `action' $-\frac{1}{2g} \sin(x)^2$ in \eqref{eq:0dPerodicPotPartFun}  has two saddle points: a `perturbative' saddle point $x=0$ with action $S_0=0$, and a `non-perturbative' saddle-point at $x=\pi/2$ with action $S_1 = 1/(2g)$.  It is natural --- and correct --- to expect that the non-perturbative saddle point needs to be taken into account to resolve the issues with naive perturbation theory that we highlighted above.  The question is how this is to be done systematically.   When $\mathcal{Z}(g)$ is evaluated perturbatively around each saddle point, the contributions can be expressed as power series weighed by exponentials of the saddle-point actions.  It is then tempting to guess that that the contributions of fluctuations of the  non-perturbative saddle must simply be \emph{added} to the perturbative series: 
\begin{align}
\mathcal{Z}(g) \overset{?}{=} e^{-S_0} \sum_{k=0}^{\infty} p_{k,0} g^k + e^{-S_1} \sum_{k=0}^{\infty} p_{k,1} g^k\,.
\label{eq:ZPerturbativeNaive}
\end{align}
However, \eqref{eq:ZPerturbativeNaive} is wrong. While both terms on the right-hand-side do indeed appear in the correct answer, \eqref{eq:ZPerturbativeNaive} does not correctly reflect the analytic continuation properties of the problem.  The correct approach is to use the technology of resurgent transcendental series (or simply \textit{transseries}), which we now explain.  

To compute $p_{n,0}$ we work with the second rescaled integral in \eqref{eq:0dPerodicPotPartFun}, which is canonically normalized for working out the perturbative expansion. 
Evaluating \eqref{eq:0dPerodicPotPartFun} term by term in the $g \to 0$ limit gives
\begin{align}
\mathcal{Z}(g) \big|_{y=0} &=  \int\limits_{-\pi/(2\sqrt{g})}^{+\pi/(2\sqrt{g})} dy\left[ e^{-\frac{y^2}{2}}+ \frac{1}{6} g e^{-\frac{y^2}{2}} y^4+\cdots \right]
\nonumber\\
&= \sqrt{2\pi} \left(1+\frac{g}{2} +\cdots \right) \nonumber \\
&= \sqrt{2\pi} \sum_{k=0}^{\infty} \frac{\Gamma(k+1/2)^2 2^k}{\Gamma(k+1) \Gamma(1/2)^2} g^k \equiv e^{-S_0}\sum_{k=0}^{\infty} p_{k,0} g^k \equiv e^{-S_0} \Phi_0(g)\,.
\end{align}
Note that the series coefficients blow up factorially, meaning that $\Phi_0(g)$ is an asymptotic series.  Similar manipulations also yield the contribution from the non-perturbative saddle-point:
\begin{align}
 e^{-1/(2g)}\sqrt{2\pi} \sum_{k=0}^{\infty} \frac{(-1)^k\Gamma(k+1/2)^2 2^k}{\Gamma(k+1) \Gamma(1/2)^2} g^k 
&= e^{-1/(2g)}\sum_{k=0}^{\infty} p_{k,1} g^k \equiv e^{-S_1} \Phi_1(g)\,
\end{align}
and we observe that $p_{k,1} = (-1)^k p_{k,0}$.  This is also a divergent series.  So both terms on the right-hand side of \eqref{eq:ZPerturbativeNaive} are asymptotic series. 

For Gevrey order-1 series (i.e, $p_k \sim  k!$),  
 one can make sense of asymptotic series by Borel summation.  The Borel transforms of $\Phi_0(g)$ and $\Phi_1(g)$ are 
\begin{align}
B \Phi_0(t) &= \sqrt{2\pi} \sum_{k=0}^{\infty}  \frac{p_{k,0}}{k!}  t^k\,, \\
B \Phi_1(t) &= \sqrt{2\pi} \sum_{k=0}^{\infty} \frac{p_{k,1}}{k!} t^k\,.
\end{align}
The Borel sum of a formal power series $\Phi(g)$, when it exists, is given by an integral of the analytic continuation $\widetilde{B \Phi}(t)$ of $B \Phi(t)$ from $t=0$ to $t=+\infty$ along the real axis:
\begin{align}
\mathcal{S}\Phi(g) =  \frac{1}{g} \int_{0}^{+\infty} dt \, e^{-t/g} \widetilde{B \Phi}(t)\,.
\end{align}
When $\mathcal{S}\Phi(g)$ exists, it is a function with the same asymptotic expansion as $\Phi(g)$ by construction, but it is well-defined in some finite neighborhood of $g=0$.  In this sense, $\mathcal{S}\Phi(g)$ is the resummation of $\Phi(g)$.   

This 19th century machinery does not fully apply to our situation, however.  One can verify that $\widetilde{B \Phi}_1(t)$ has no singularities on $\mathbb{R}^{+}$, so the standard Borel sum of $\Phi_1(g)$ exists. However, $\widetilde{B \Phi}_0(t)$ has a branch-cut singularity on the $\mathrm{arg}(g) =: \theta=0$ line originating at $t=1/2$.  We also note that $\widetilde{B \Phi}_1(t)$ has a branch-cut singularity along  $\mathbb{R}^{-}$ originating at $t=-1/2$.  The $\theta = 0$ and $\theta = \pi$ rays are called \emph{Stokes rays}, while the $0<\theta < \pi$ and $-\pi < \theta<0$ half-planes are called \emph{Stokes chambers}.  The standard Borel sum involves an integral along a Stokes ray of our system of interest, and as a result, the standard Borel sum of $\Phi_0(g)$ does not exist.  Our problem is a classic example of a \emph{non-Borel summable} series.   

Note, however, that we can also define a generalized Borel sum associated with integrals along other rays in the complex $t$ plane
\begin{align}
\mathcal{S}_{\theta} \Phi_{0,1}(g) =  \frac{1}{g} \int_{0}^{+\infty e^{i\theta}} dt \, e^{-t/g} \widetilde{B \Phi}_{0,1}(t)\,.
\label{eq:GeneralizedBorel}
\end{align}
This integral exists if $\theta \neq 0$ and $\theta \neq \pi$ for $\Phi_0$ and $\Phi_1$ respectively.  But 
for $\Phi_0$,  the result of this generalized Borel summation is \emph{complex-valued}  
for $\theta=0^{\pm}$, and \emph{ambiguous} at $\theta=0$ (for real $g$) as the $\theta=0$ limit  of the integration  from above and below do not agree.  
Since we started with a problem defined by a manifestly \emph{real} integral for $g \in \mathbb{R}^{+}$, getting a complex-valued and ambiguous result may not seem like progress.  Yet it turns out that understanding the relation of \eqref{eq:GeneralizedBorel} to \eqref{eq:0dPerodicPotPartFun} is an essential step.  

To see this, we introduce some terminology from resurgence theory.  We will call expressions of the form
\begin{align}
\mathcal{Z}(g, \sigma_0, \sigma_1) = \sigma_0 e^{-S_0} \sum_{k=0}^{\infty} p_{k,0} g^k + \sigma_1 e^{-S_1} \sum_{k=0}^{\infty} p_{k,1} g^k\,,
\label{eq:Ztranseries}
\end{align}
\emph{transseries}, and will refer to $\sigma_{0,1}$ as \emph{transseries parameters}.  The two distinct power series (weighted by exponentials) represent contributions from the two distinct integration cycles 
in a complexified version of \eqref{eq:0dPerodicPotPartFun} where the variable $x$ is complex.
These two cycles are associated with the steepest descent paths of the two critical points. 
 The appearance of $\sigma_{0,1}$ parametrizes the generalized partition functions $\mathcal{Z}(g, \sigma_0, \sigma_1)$ which can be defined using various linear combinations of the convergent integration cycles of the complexified integral.   The transseries parameters are  constant 
 in any given Stokes chamber, but their values jump across Stokes rays, i.e, they are {\it piece-wise constant}.   These jumps are the key feature which explains why the complex-valued generalized Borel sums are so useful in understanding our original real integral. We refer the interested reader to e.g. \cite{Dunne:2012zk,Cherman:2014ofa} for more details about these matters. 

In the present case, we are in the fortunate position of knowing exact expressions for the analytic continuations of the Borel transforms
\begin{align}
\widetilde{B\Phi}_0(t) &= \sqrt{2\pi} {}_{2}F_{1}\left(\frac{1}{2},\frac{1}{2},1; 2t\right)\,, \\
\widetilde{B\Phi}_1(t) &=\sqrt{2\pi}  {}_{2}F_{1}\left(\frac{1}{2},\frac{1}{2},1; - 2t \right)\,.
\end{align}
This can be verified by comparing the above formulae with the power series expressions for the hypergeometric function\footnote{See e.g. $\textrm{http://dlmf.nist.gov/15.2\#E1}. $}
\begin{align}
{}_{2}F_1(a, b; c; z) = \frac{\Gamma(c)}{\Gamma(a) \Gamma(b)} \sum_{n = 0}^{\infty} \frac{\Gamma(a+n)\Gamma(b+n)}{\Gamma(c+n) \Gamma(n+1)} z^{n}.
\end{align} 
Having such exact analytic continuations is a rare luxury, and we emphasize that it is not necessary to deduce the monodromy properties which will be crucial below.  What is necessary is to have information on the large-order asymptotics of $\Phi_{0,1}(g)$, and such asymptotic information is usually much easier to find than an exact expression for $\widetilde{B\Phi}_{0,1}$.

The Borel sum of $\Phi_0$ along $\theta=0$ does not exist, but it \emph{does} exist for $\theta = 0^{\pm}$.  The difference between these directional (lateral) Borel resummations is imaginary:
\begin{align}
(\mathcal{S}_{0^+} - \mathcal{S}_{0^-}) \Phi_0(g)  &=\lim_{\epsilon \to 0} \frac{\sqrt{2\pi} }{g} \int_{1/2}^{\infty} dt \, e^{-t/g}
\left[{}_{2}F_{1}\left(\frac{1}{2},\frac{1}{2},1; 2t + i\epsilon \right)  - {}_{2}F_{1}\left(\frac{1}{2},\frac{1}{2},1; 2t - i\epsilon \right) \right] \nonumber \\
&=\frac{2i \sqrt{2\pi} }{g} \int_{1/2}^{\infty} dt \, e^{-t/g} {}_{2}F_{1}\left(\frac{1}{2},\frac{1}{2},1; 1-2t\right) \nonumber \\
&=\frac{2 i \sqrt{2\pi} }{g} e^{-1/(2g)} \int_{0}^{\infty} dt\,e^{-t/g}
{}_{2}F_{1}\left(\frac{1}{2},\frac{1}{2},1; 1-2t\right) \nonumber\\
&= 2 i e^{-1/(2g)}\mathcal{S}_0 \Phi_{1}(g)\,.
\label{eq:HankelContourJump}
\end{align}
Here we used the known discontinuity properties of hypergeometric functions, but as remarked above, all that is actually necessary to establish \eqref{eq:HankelContourJump} is a knowledge of the asymptotic form of $p_{n,0}$. 

Demanding that the transseries representation of $\mathcal{Z}(g)$ be ambiguity-free for any complex $g$ means that (with $\sigma_0 = 1$) we must take $\sigma_1 = - i$ for $\pi>\arg \theta > 0$ and $\sigma_1 = + i$ for $-\pi<\arg \theta <0$.   Hence the imaginary ambiguity in the resummation of $\Phi_0$ at $\theta=0$ is given by the summation of $\Phi_1$.  Note that this means that the large-order behavior of $\Phi_0(g)$, which determines the ambiguities in its resummation near $\theta = 0$, is controlled by the low-order behavior of $\Phi_1(g)$.   This is the origin of the term `resurgence':  information about other saddle points pops up in a coded form (resurges) in the expansion around any given saddle point.     

\begin{figure*}[t]
  \centering
\includegraphics[width=0.95\textwidth]{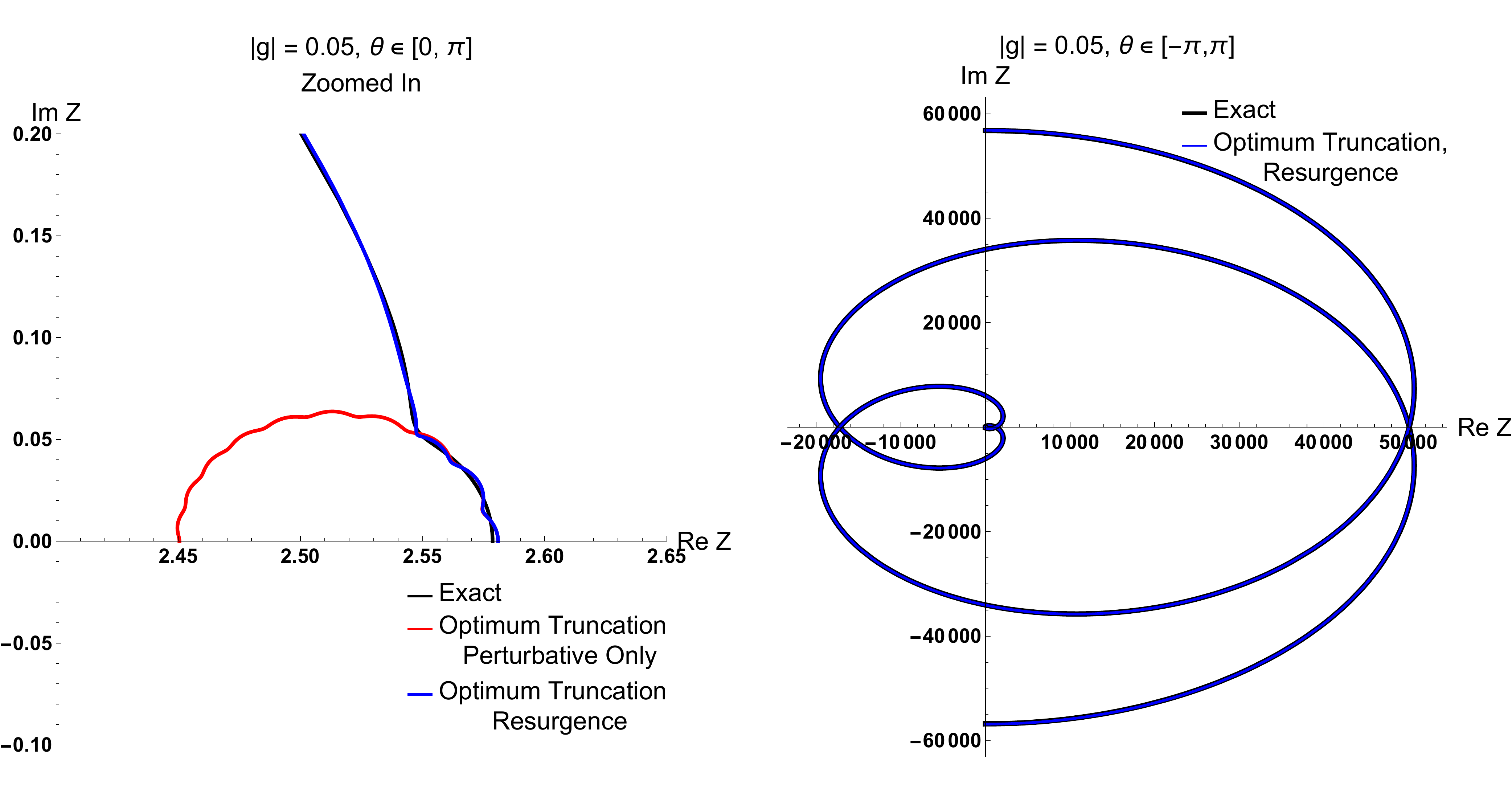}
  \caption{(Color Online.) Behavior of $\mathcal{Z}$ as a function of $\theta = \arg g$ with fixed $|g| \ll 1$.  The figure on the left illustrates that the optimally-truncated perturbative series alone only gives a good approximation to the exact answer for small $\theta$.   In contrast, \eqref{eq:ResurgenceExpressionTheta} (also treated with optimal truncation) gives an excellent approximation to the exact result for all values of $\theta$, as illustrated in the right figure.}
  \label{fig:FiniteTheta}
\end{figure*}

A very important consequence of \eqref{eq:HankelContourJump} is that the transseries representation of $\mathcal{Z}(g)$ is ambiguity-free even at $\theta = 0^{\pm}$ thanks to jumps in $\sigma_{0,1}$ as $\theta$ crosses the $\theta = 0$ Stokes line.  In particular, our original real partition function can be written as 
\begin{align}
\label{eq:ResurgenceExpression}
\mathcal{Z}(g) |_{ \theta  = 0}&= \mathcal{S}_{0^{\pm}} \Phi_{0}(g) \mp i e^{-\frac{1}{2g}} \mathcal{S}_{0^{\pm}} \Phi_{1}(g) \\
&= \mathrm{Re} \mathcal{S}_{0
} \Phi_{0}(g) \nonumber. 
\end{align}
which is real and unambiguous. 
Right at $\theta = 0$, the only role of the nonperturbative saddle point is simply to cancel the imaginary part of $\mathcal{S}\Phi_{0} (g)$.  If one works with an integral which is real in the first place, and $g \ll 1$, it is tempting to think that any exponentially small imaginary part should be dropped simply on intuitive grounds.  From this perspective one may wonder whether the argument leading to \eqref{eq:ResurgenceExpression} may be a complicated derivation of an intuitively obvious fact.  This is not so.  First, as we explore in the next section, \eqref{eq:ResurgenceExpression} has major consequences at strong coupling, where the non-perturbative contributions are not small.  Second, it has major consequences already at weak coupling.  The issue is that as soon as we allow $g$ to have a complex phase, the non-perturbative contributions again cease being small, and \eqref{eq:ResurgenceExpression} becomes essential for reconstructing $\mathcal{Z}(g)$.   In particular, if $\arg{g} = \theta \in [-\pi, \pi]$, the resurgence relation between $\mathcal{Z}(g)$ and its transseries representation is
\begin{align}
\label{eq:ResurgenceExpressionTheta}
\mathcal{Z}(g) = 
\begin{cases}
    \mathcal{S}_{\theta} \Phi_{0}(g) - i e^{-\frac{1}{2g}} \mathcal{S}_{\theta} \Phi_{1}(g), & \theta \in (0, \pi) \\
    \mathcal{S}_{\theta} \Phi_{0}(g) + i e^{-\frac{1}{2g}} \mathcal{S}_{\theta} \Phi_{1}(g), & \theta \in (-\pi,0)
\end{cases}
\end{align}
This expression can be viewed as the corrected version of \eqref{eq:ZPerturbativeNaive}. 

The left-hand side of Fig.~\ref{fig:FiniteTheta} illustrates the behavior of $\mathcal{Z}(g)$ versus its \emph{full} transseries representation \eqref{eq:ResurgenceExpressionTheta}, treated within the approximation of optimal truncation\footnote{\emph{Optimal truncation} for asymptotic series means summing the series to order $n_{*} = 1/g$, where it is assumed that $g \ll 1$.  Past $n_{*}$, the terms in an asymptotic series start becoming larger rather than smaller. This truncation condition is termed `optimal' because it minimizes the error due to the use of the perturbative expression for the given value of $g \ll 1$. 
The error in optimal truncation is  of a  non-perturbative form $e^{-A/g}$, where  $A$ is the relative action with respect to nearest other saddle.}
 with fixed $|g| \ll 1$ and varying $\theta$.  In fact, to the level of accuracy that is visible in the figures by the naked eye with our chosen value of $|g|$, we could have simply used a fixed-order truncation approximation in all of the perturbative series.  Note the close agreement with the exact result even for $|g| = 0.05$  obtained with the aid of the above resurgence formulae.  This can be contrasted with the very poor accuracy of taking into account \emph{only} the optimally-truncated perturbative series as soon as $\theta$ deviates from $0$ appreciably, as was already illustrated in Fig.~\ref{fig:NaivePerturbationTheory}.

Expressions like \eqref{eq:ResurgenceExpression} are also very useful from another point of view.  For instance, suppose we originally knew nothing about the existence of the non-perturbative saddle point, but we did know the perturbative contribution to $\mathcal{Z}(g)$.  Then demanding that $\mathcal{Z}(g)$ be ambiguity free across the $\theta = 0$ Stokes line would tell us that (a) the nonperturbative contribution must exist and (b) its contribution is completely determined by the discontinuity structure of $\mathcal{S} \Phi_{0}(g)$.  A similar argument allows one to reconstruct the perturbative contribution from the discontinuity structure of $\mathcal{S} \Phi_{1}(g)$ Stokes line.  

For future use, we note that the the above discussion in terms of the monodromy properties of the integration cycles of $\CZ(g)$ can be summarized as follows. The Borel-sum ambiguities of $\Phi_{0,1}$ across the $\theta=0, \pi$ Stokes rays imply  the following  upper-triangular Stokes matrix at $\theta=0$ for the $\mathcal{J}_{0,1}$ steepest-descent integration cycles (so-called Lefshetz thimbles) associated the perturbative and non-perturbative contributions (see \cite{Cherman:2014ofa} for more details):
\begin{align}
 \begin{pmatrix}
\mathcal{J}_0  \\
\mathcal{J}_1
\end{pmatrix} \to 
 \begin{pmatrix}
1 & -2 \\
0 & 1
\end{pmatrix} 
\cdot
 \begin{pmatrix}
\mathcal{J}_0  \\
\mathcal{J}_1
\end{pmatrix} , \qquad \theta = 0
\end{align}
At $\theta = \pi$,  the Stokes matrix is lower triangular and is given by 
\begin{align}
 \begin{pmatrix}
\mathcal{J}_0  \\
\mathcal{J}_1
\end{pmatrix} \to 
 \begin{pmatrix}
1 & 0 \\
2 & 1
\end{pmatrix} 
\cdot
 \begin{pmatrix}
\mathcal{J}_0  \\
\mathcal{J}_1
\end{pmatrix} , \qquad \theta = \pi .
\end{align}
So in the basis of the steepest-descent curves, which is the most natural one when $g \ll 1$, the (formal) monodromy matrix can be expressed as the following product of the above Stokes matrices
\begin{align}
M_{g=0} =  \begin{pmatrix}
1 &  0\\
2 & 1
\end{pmatrix}  
\cdot
 \begin{pmatrix}
1 & -2 \\
0 & 1
\end{pmatrix} 
=  \begin{pmatrix}
1 & -2 \\
2 & -3
\end{pmatrix}\,.
\label{eq:Monodromyg0PerCase}
\end{align}

\subsection{Strong Coupling Results from Weak Coupling Data}\label{Sec:StrongTrig}

\begin{figure*}[t]
  \centering
\includegraphics[width=0.48\textwidth]{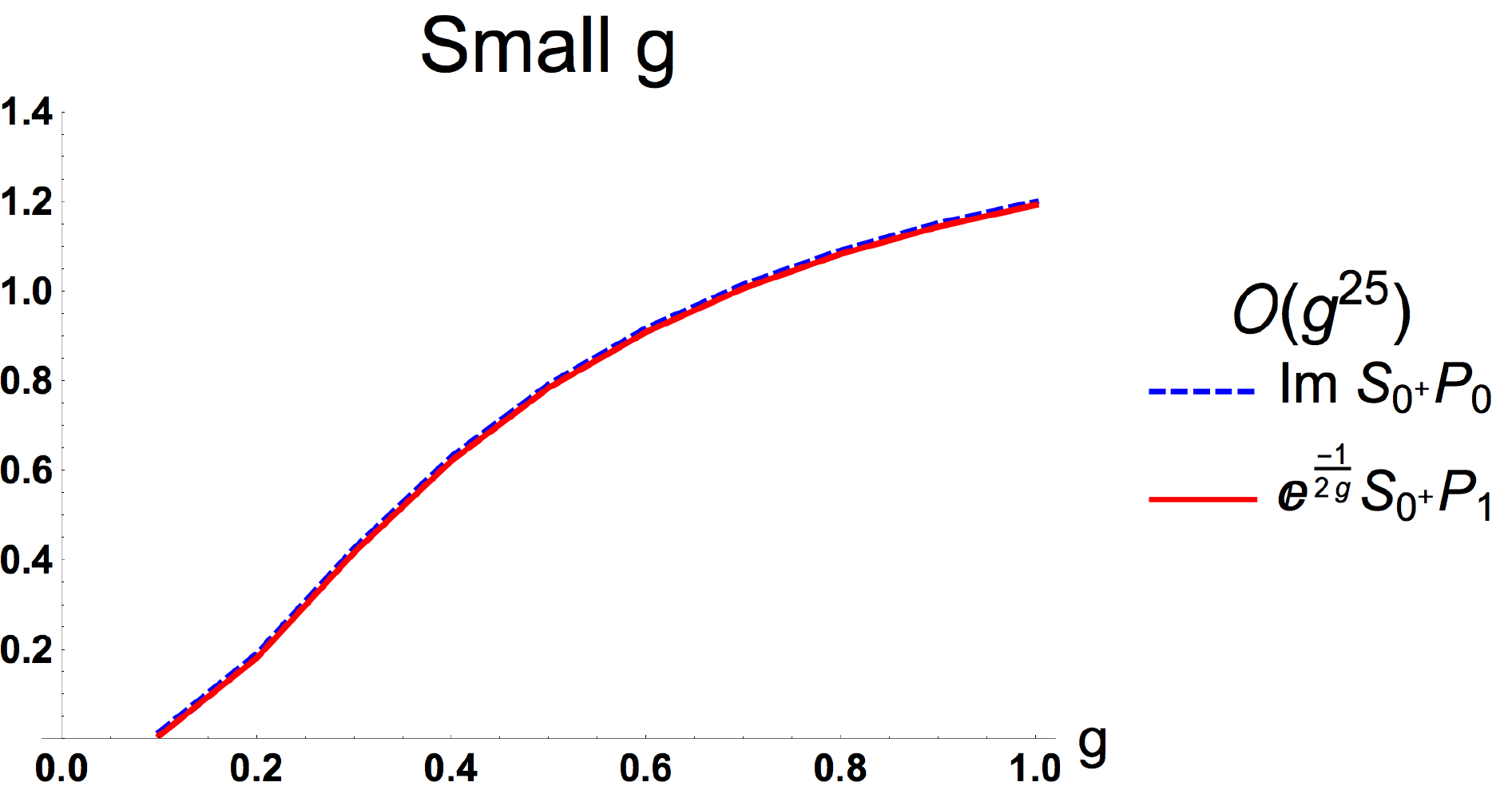}
\includegraphics[width=0.48\textwidth]{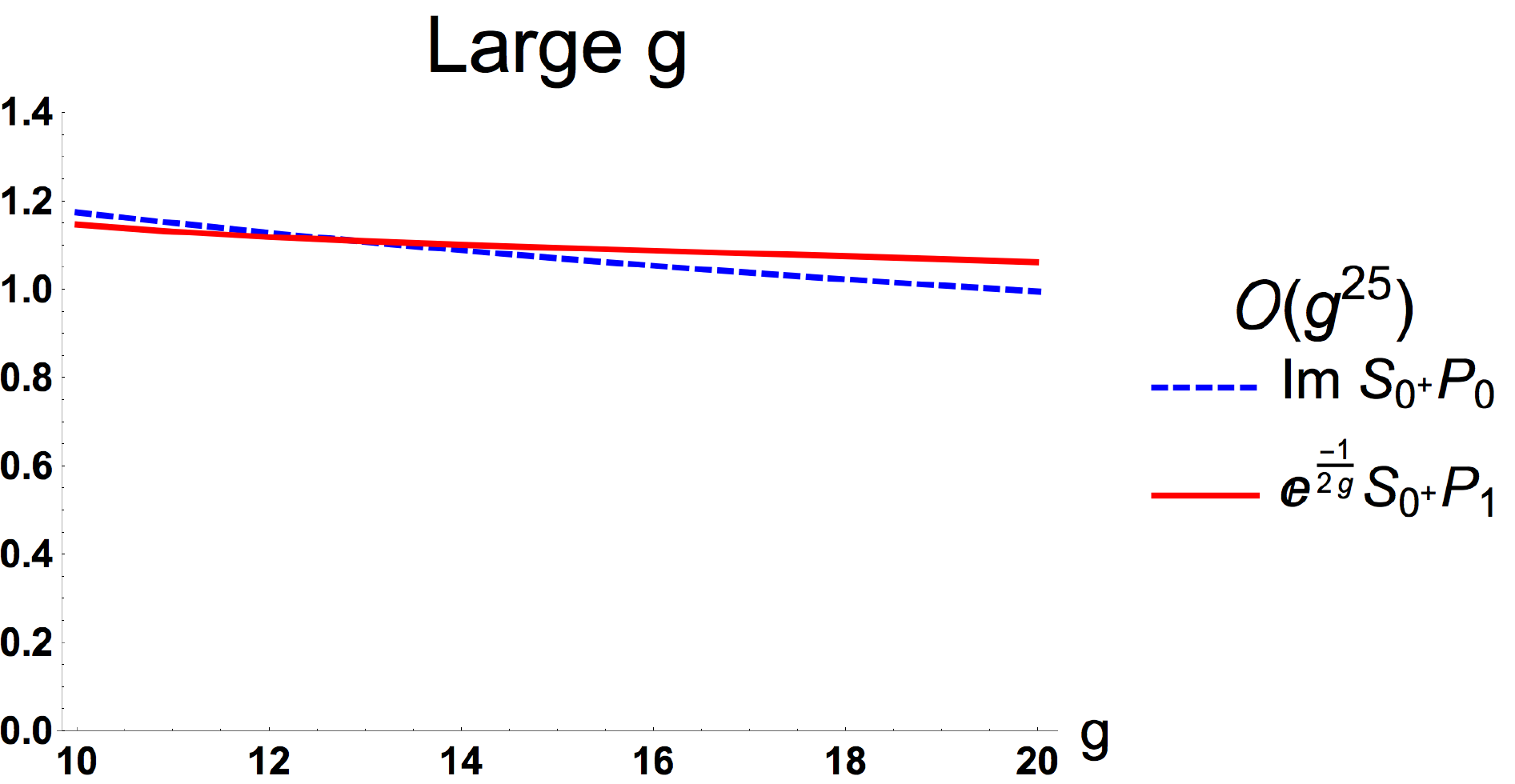}
  \caption{(Color Online.) Behavior of $\mathrm{Im}\, \mathcal{S}_{0^{\pm}} \Phi_{P}(g)$ and $e^{-\frac{1}{2g}} \mathcal{S}_{0^{\pm}} \Phi_{NP}(g)$ evaluated using $N=12$, corresponding to taking terms up to $g^{25}$ in the weak-coupling expansion, with $\theta = 10^{-3} \pi$, at small (left) and large (right) $g$.  In the limit $\theta \to 0^{+}$ and $N\to \infty$ the two curves come to lie on top of each, as demanded by \eqref{eq:ResurgenceExpression}.}
  \label{fig:monodromyCheck}
\end{figure*}

\begin{figure*}[th]
  \centering
\includegraphics[width=.85\textwidth]{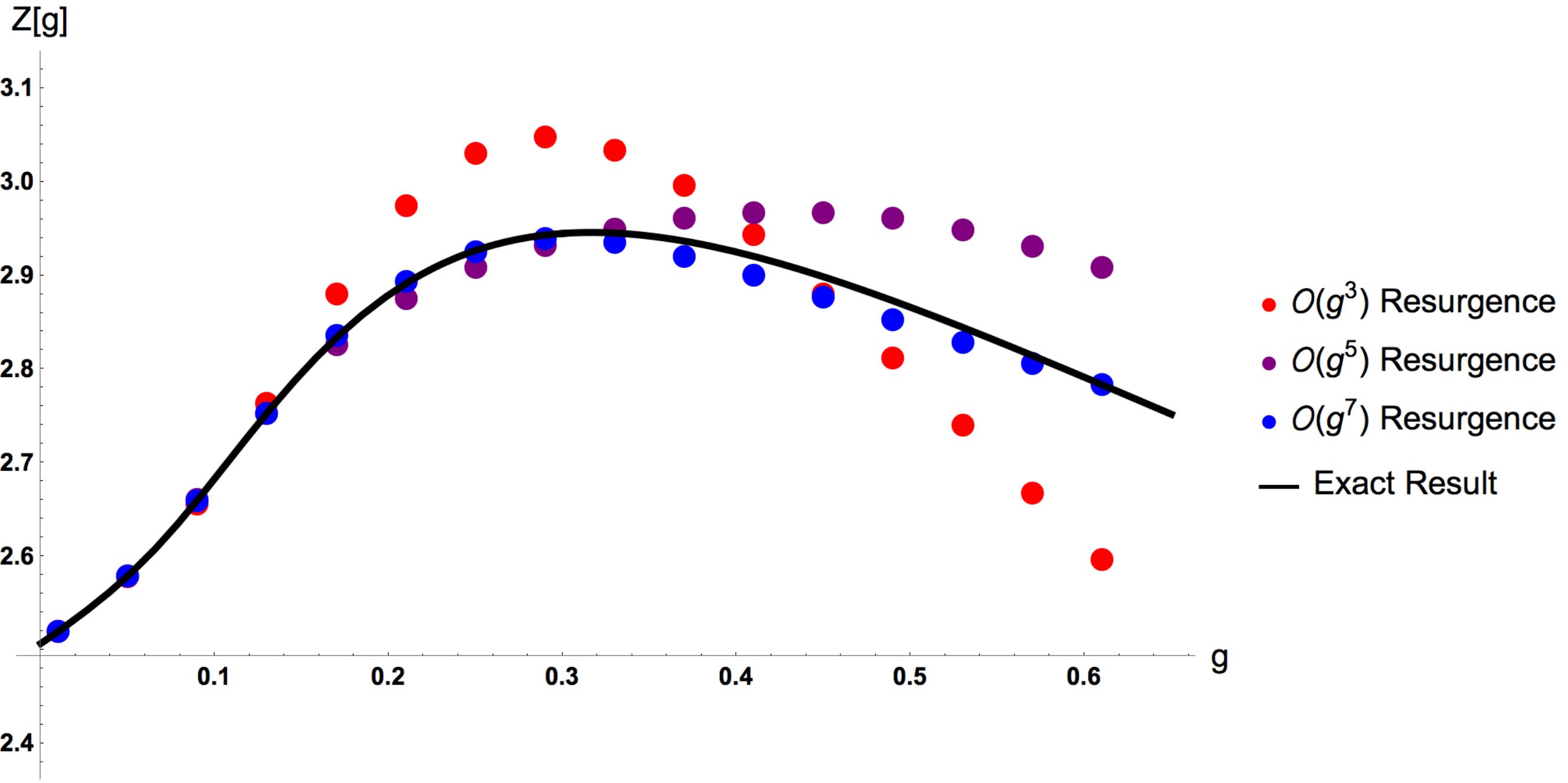}
\includegraphics[width=.85\textwidth]{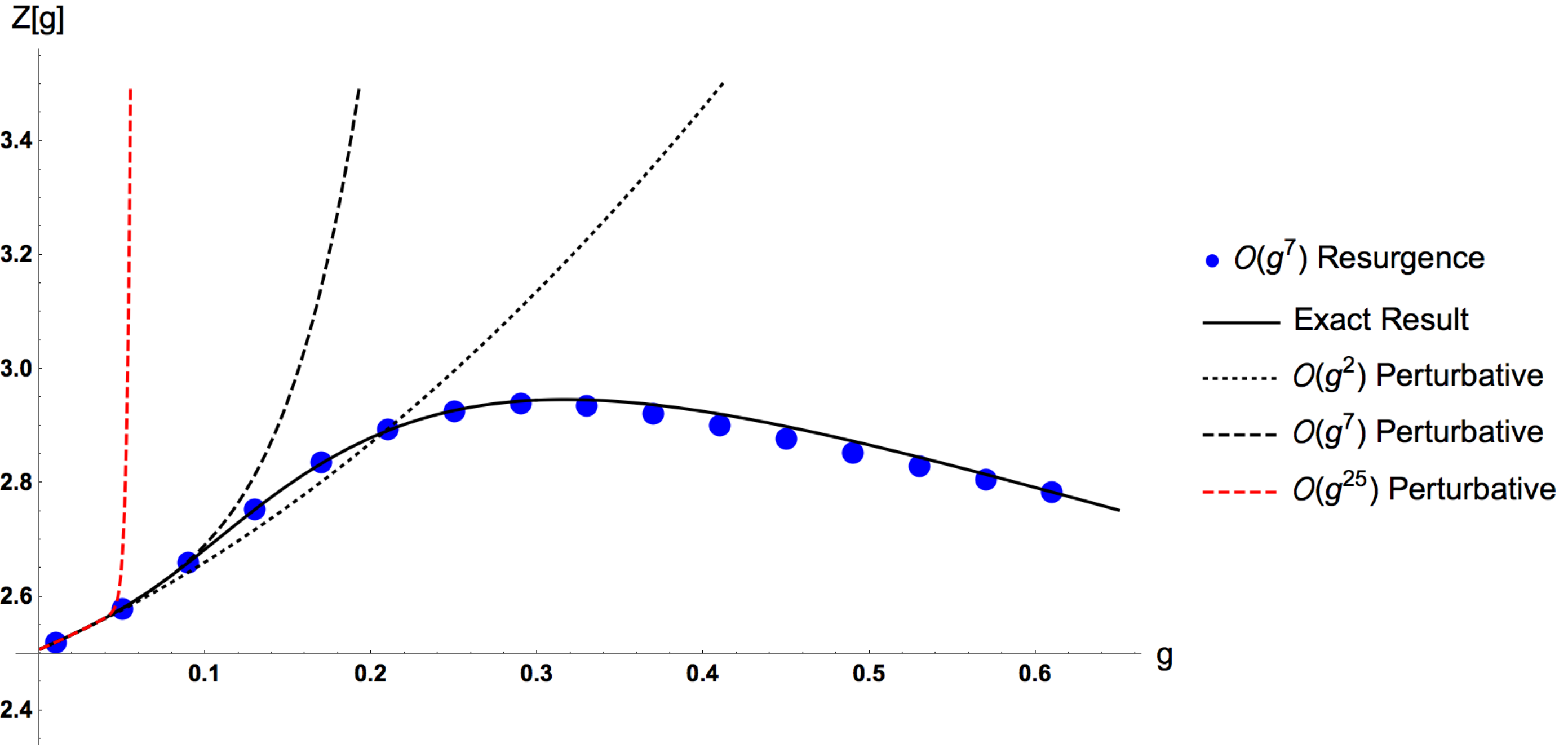}
  \caption{(Color Online.) We illustrate the power of Borel-\`Ecalle summation at low values of $g$.  The top figure shows the convergence of the Borel-\`Ecalle sums to the exact expression for $\mathcal{Z}(g)$ with perturbative data through $\mathcal{O}(g^3), \mathcal{O}(g^5)$ and $\mathcal{O}(g^7)$.  The bottom figure compares the accuracy of Borel-\`Ecalle summation to naive partial sums of the perturbative series at various orders.  }
  \label{fig:smallGResurgencePlot}
\end{figure*}

\begin{figure*}[th]
  \centering
\includegraphics[width=.85\textwidth]{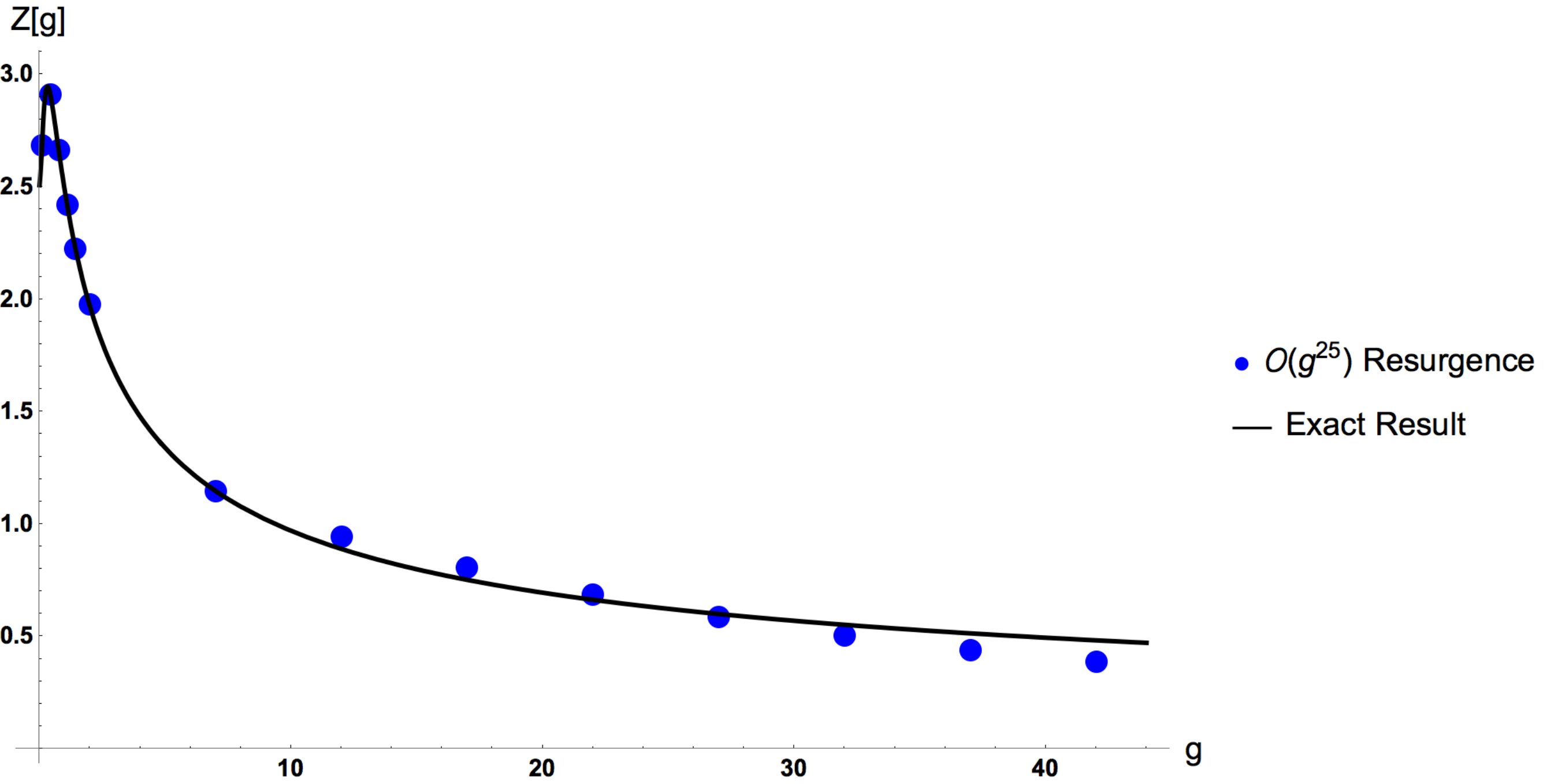}
  \caption{(Color Online.) Comparison of the predictions from Borel-\`Ecalle summation (with $\theta = 10^{-3} \pi \ll 1$) of the data contained in terms up to $\mathcal{O}(g^{25})$  $g^{25}$ in the weak-coupling expansion (blue dots), to the exact result at strong coupling (black curve).}
  \label{fig:strongVsWeakPlot}
\end{figure*}

\begin{figure*}[th]
  \centering
\includegraphics[width=.85\textwidth]{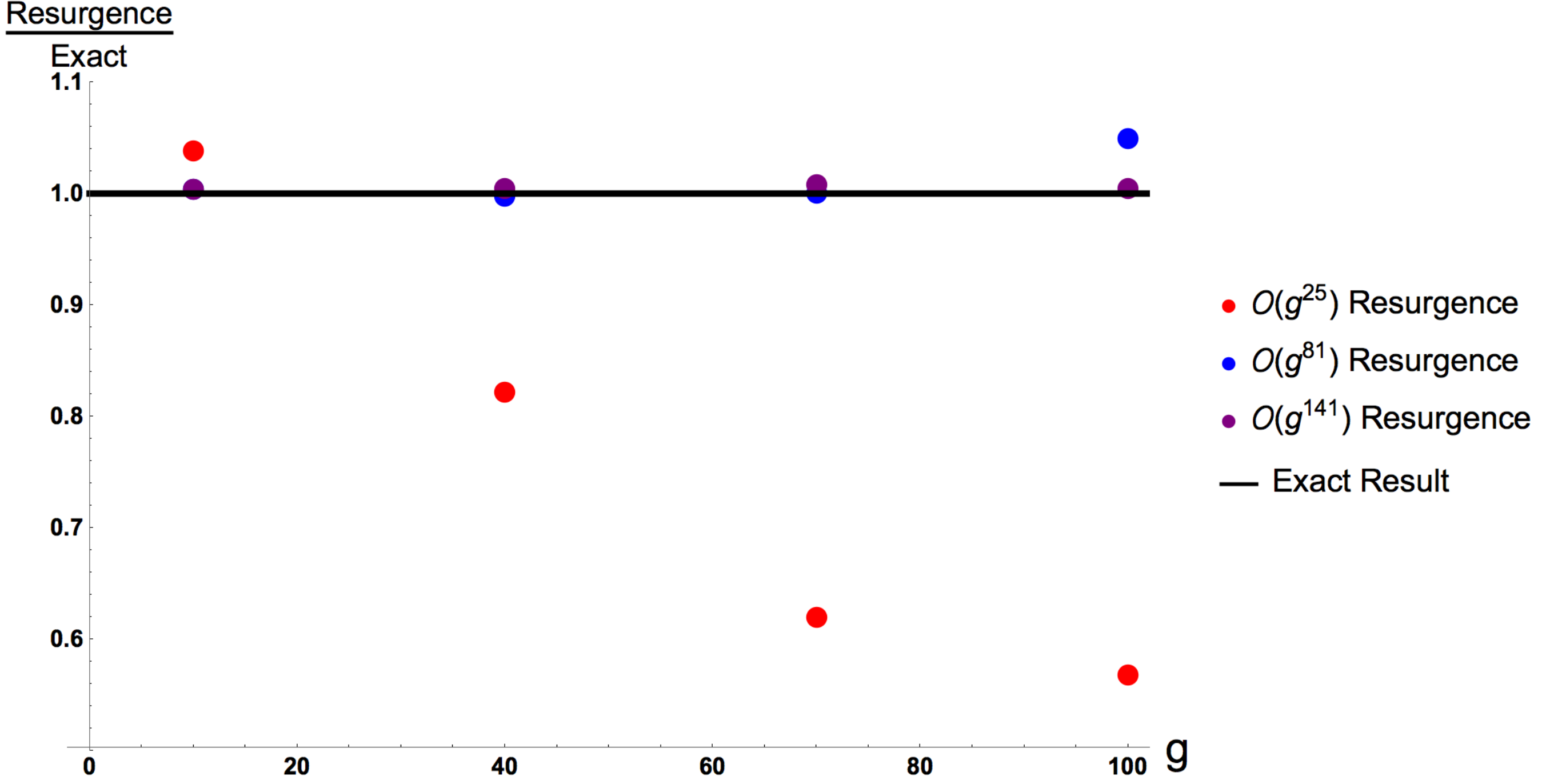}
  \caption{(Color Online.) One must use data from higher orders in the small $g$ expansion to get accurate strong-coupling approximations from Borel-\`Ecalle summation of weak-coupling data as $g$ is increased.  The plot shows the result of Borel-\`Ecalle summation normalized to the exact answers as a function of $g$ and the number of terms from the weak-coupling expansion used in the calculation. }
  \label{fig:convergencePlot}
\end{figure*}

In this simple example we are in the luxurious position of knowing the expansion of $\mathcal{Z}(g)$ at small $g$, large $g$, and indeed we know $\mathcal{Z}(g)$ exactly at any $g$.  In more complicated examples one is usually only in possession of a weak-coupling asymptotic series expansion for $\mathcal{Z}(g)$.   Here we show that given  the monodromy information from resurgence theory, which is derivable from the weak coupling expansion,  one can extract strong-coupling results from the weak-coupling data.

We have seen that the cancellation of ambiguities in the weak-coupling transseries representation of $\mathcal{Z}(g)$ implied by \eqref{eq:HankelContourJump} is summarized by \eqref{eq:ResurgenceExpression}, which we repeat here for convenience:
\begin{align}
\mathcal{Z}(g) &= \mathcal{S}_{0^{\pm}} \Phi_{0}(g) \mp i e^{-\frac{1}{2g}} \mathcal{S}_{0^{\pm}} \Phi_{1}(g)\notag \\
&= \mathrm{Re} \mathcal{S}_{0
} \Phi_{0}(g)\,.  
\end{align}
This is a highly non-trivial piece of `global' information about the relationship between the local data in the Borel sum of the small-$g$ power series $\Phi(g)$ and $\mathcal{Z}(g)$.  When $g \ll 1$, the imaginary part of  $ \mathcal{S}_{0^{\pm}} \Phi_{0}(g)$ is exponentially small, and is negligible compared to $g^n$ for any fixed $n$, so at weak coupling one might be tempted to discard $\mathrm{Im} \mathcal{S}_{0^{\pm}} \Phi_{0}(g)$ on an ad-hoc basis.  But in this simple example, resurgence theory tells us that $\mathrm{Im} \mathcal{S}_{0^{\pm}} \Phi_{P0}(g) $ cancels against the contribution of the nonperturbative saddle point for \emph{any} $g$ if $\arg g \to 0$.  In particular, resurgence theory tells that this continues to be true even when $g$ is large, and $e^{-1/(2g)}$ is no longer small, so that the NP contribution cannot be dropped on hand-waving grounds!

This insight from resurgence theory is enough to compute the strong coupling behavior of $\mathcal{Z}(g)$ from its weak coupling expansion via \eqref{eq:ResurgenceExpression}.  Given a \emph{finite-order} approximation to the formal power series $\Phi_0(g)$ 
\begin{align}
\Phi^{(2N+1)}_{0}(g) = \sqrt{2\pi} \sum_{n=1}^{2N+1} a_n g^n\,,
\end{align}
a finite-order approximate to the Borel transform is
\begin{align}
B\Phi^{(2N+1)}_{0}(t) = \sqrt{2\pi} \sum_{n=1}^{2N+1} \frac{a_n}{n!} t^n\,.
\label{eq:BorelApprox}
\end{align}
To compute  $\mathrm{Re} \mathcal{S}_{0^{\pm}} \Phi_{0}(g)$, we must construct an approximation for the analytic continuation $\widetilde{B \Phi}_0(t)$ of $B \Phi_0(t)$.  The original power series \eqref{eq:BorelApprox} itself is a very poor approximation to $\widetilde{B \Phi}_0(t)$ in the cases of interest, since (being a power series) it converges in a circle in the $t$ plane, while we expect $\widetilde{B \Phi}_0(t)$ to have cuts in general.   A much more efficient approach to computing approximations to $\widetilde{B \Phi}_0(t)$ is to use a series of functions of $t$ which encode the same information as \eqref{eq:BorelApprox}, but converge in a cut plane rather than within a disk.    One example of such a series is furnished by the Pad\'e approximants.    We define a Pad\'e approximant to $B \Phi^{(2N+1)}_{0}(t)$ to be  a rational function\footnote{The objects we are using are called \emph{diagonal Pad\'e approximants}.}
\begin{align}
R^{(2N+1)}(t) = \frac{\sum_{k=0}^{N} c_k t^k}{1+\sum_{m=1}^{N} d_m t^m}
\label{eq:Pade}
\end{align}
with a Taylor series which agrees with $B \Phi^{(2N+1)}_{P}(t)$ through order $2N+1$.  We will refer to the summation scheme based on \eqref{eq:ResurgenceExpressionTheta}, \eqref{eq:BorelApprox}, and \eqref{eq:Pade} as \emph{Borel-\`Ecalle summation}.  This name is chosen because the key to the summation procedure is its systematic incorporation of the global analytic structure of the system via resurgence theory, as reflected in \eqref{eq:ResurgenceExpressionTheta}.  The use of Pad\'e approximants in the numerical evaluation of generalized Borel sums is incidental to the procedure, and any other analytic continuation of the power series which converges within cut planes could also be used\footnote{Convergence in a cut plane is important because we expect the Borel transforms to be defined on a plane with cuts for the class of functions to which resurgence theory applies, which are the so-called `endlessly continuable' functions (see e.g. \cite{Dorigoni:2014hea}).}.

Using $R^{(2N+1)}(t)$ as an approximation to  $\widetilde{B\Phi}_0(t)$ we can now compute an approximation to $\mathrm{Re} \mathcal{S}_{0^{\pm}} \Phi_{0}(g)$ using the first $2N+1$ coefficients from the small-$g$ expansion of $\mathcal{Z}(g)$, with the Borel sum integral evaluated numerically with $g \to g e^{i \theta}$ with $\theta \ll 1$.~\footnote{We take $\theta = 10^{-3} \times \pi$ since decreasing $\theta$ further changes the results by a negligible amount.}   The approximation is improvable by increasing $N$.

In Fig.~\ref{fig:monodromyCheck} we show that,   the cancellation of the ambiguity \eqref{eq:ResurgenceExpression} holds to high accuracy both at small $g$ and large $g$ when the quantities on the right hand side of the first line are approximated using Borel-\`Ecalle summation.

Next we show that \eqref{eq:ResurgenceExpression} used together with the weak-coupling expansion can be used to obtain highly accurate approximations to $\mathcal{Z}(g)$. First, we work with $g \lesssim 1$.  In Fig.~\ref{fig:smallGResurgencePlot} show the convergence of Borel-\`Ecalle approximants using weak coupling information through $\mathcal{O}(g^3), \mathcal{O}(g^5)$ and $\mathcal{O}(g^7)$ to the exact values of $\mathcal{Z}(g)$ for $g \lesssim 1$.   Fig.~\ref{fig:smallGResurgencePlot} also shows that Borel-\`Ecalle summation gives a far better approximation to the value of $\mathcal{Z}(g)$ than a partial sum of the perturbative series away from infinitesimal values of $g$.  This is a reflection of the fact that the perturbative series is asymptotic, and has a zero radius of convergence.  Nevertheless, it contains very valuable data about the behavior of $\mathcal{Z}(g)$ in a  coded form, and Borel-\`Ecalle summation decodes this data.

Having seen that Borel-\`Ecalle  summation yields excellent approximations to $\mathcal{Z}(g)$ for $g \lesssim 1$, let us examine what happens for larger $g$.  In Fig.~\ref{fig:strongVsWeakPlot} we plot the exact expression for $\mathcal{Z}(g)$ versus the Borel-\`Ecalle approximations using \eqref{eq:ResurgenceExpression} and perturbative data through $\mathcal{O}(g^{25})$ (equivalently, $N=12$).  This relatively low-order approximation works very well all the way out to $g \sim 40$.  Note also that the Borel-\`Ecalle resummation accurately tracks the bump in $\mathcal{Z}(g)$ at $g\sim 1$, despite the fact that naively it merely contains information from an expansion around $g=0$.

 Finally, Fig.~\ref{fig:convergencePlot} illustrates the point that by using higher-order perturbative data to build the Borel-\`Ecalle sums, we can get accurate approximations to $\mathcal{Z}(g)$ at higher and higher values of $g$.  If one wishes to extract the result at infinite coupling, this requires all orders in resurgent expansion on the weak coupling side.

\subsection{Holomorphic Analysis at Strong Coupling}
Let us now take a direct look at the strong coupling description of the problem\footnote{Analyses similar to what follows in this subsection has been discussed in the literature (see e.g. \cite{kamimoto2010,2012arXiv1202.3100V}). In the this work our emphasis is on the connection of the strong coupling monodromy analysis to resurgence theory.}. We will be using $(2\xi)^{-1}=2g$, so that \eqref{eq:0dPerodicPotPartFun} now reads
\begin{equation}
\mathcal{Z}(\xi) = \sqrt{\xi}\int\limits_{-\pi}^{+\pi} e^{-2\xi \sin^2 \phi}d\phi\,,
\label{eq:ParFuncReal}
\end{equation}
where we have doubled the integration limits to   $(-\pi,\pi)$ and divided the integral by two. 
 In what follows we shall try to reproduce \eqref{eq:HankelContourJump} without doing any explicit expansion and/or resummation. Instead, we will explore the analytic properties of $\CZ$ as a function of complexified coupling $\xi$. For our purposes it will be more convenient to work with an 
integral which is a slight variation of \eqref{eq:ParFuncReal}:
\begin{equation}
\mathcal{Z}_\ast(\xi) = \int\limits_{-\pi}^{+\pi} e^{\xi \cos \phi}d\phi = 2\pi I_0(\xi)\,, 
\label{eq:ParFuncReal1} \qquad {\rm  where}   \qquad \mathcal{Z}(\xi)=\sqrt{\xi}e^{-\xi}\mathcal{Z}_\ast(\xi)\,.
\end{equation}

In \eqref{eq:ParFuncReal1} the partition function is real and the contour of integration lies along the real $\phi$ axis. We will relax this reality condition, and only assume that some closed contour in complex $\phi$-plane is chosen. Due to the periodicity of the cosine function in the exponential we can parameterize $\cos\phi=-x$ to get
\begin{equation}
\CZ_\ast(\xi)=\int\limits_{\mathcal{C}}\frac{e^{-\xi x}}{\sqrt{(1+x)(1-x)}}dx\,,
\label{eq:ParFuncCompl}
\end{equation}
for some contour $\CC$ in the complex $x$-plane. The complex $x$-plane has two branching points at $-1$ and $1$, so there is one non-contractable cycle around the cut between these points. From the change of variable we made above, it is clear that the original integral in \eqref{eq:ParFuncReal1} can be written as an integral of the one-form
\begin{equation}
\lambda = \frac{e^{-\xi x}}{\sqrt{(1+x)(1-x)}}dx
\label{eq:LambdaForm}
\end{equation}
over this cycle, which we will call the $A$-cycle:
\begin{equation}
\CZ_\ast(\xi) = \int\limits_A \lambda\,.
\end{equation}

One may wonder if there are other nontrivial cycles on the $x$-plane. To check this, we will assume that another cycle ($B$-cycle) exists and work out the consequences of the assumption, which will then be verified self-consistently.  We will also assume that there are just two cycles on the surface.  Then Poincar\'e duality implies that there should be  two independent one-forms. Using these assumptions we can compute the integral of $\lambda$ of $B$-cycle without evaluating the integral explicitly. Indeed, let us pick some cycle $\gamma$ on the surface.  Then we can exploit the dependence on the parameter $\xi$ to generate new forms by differentiating $\lambda$. Indeed 
\begin{equation}
\frac{d}{d\xi}\int\limits_\gamma \lambda\,
\end{equation}
gives a new period integral. However, by our assumption that  there are only two independent periods, the second derivative will not give a new integral. Therefore we conclude that if our assumptions are correct, then there is a linear combination of $\CZ_\ast(\xi), \CZ_\ast'(\xi)$ and $\CZ_\ast''(\xi)$ which vanishes. Indeed, we can compute
\begin{equation}
\frac{d}{dx}\left(e^{-\xi x}\sqrt{1-x^2}\right) = \frac{\xi x^2-x-\xi}{\sqrt{1-x^2}}e^{-\xi x}\,,
\end{equation}
and observe that a closed integral of the above expression is, first, equal to zero since it is a total derivative, and, second, it is a linear combination of $\xi$ derivatives of $\CZ$. This verifies our assumptions, and we find that $\CZ_{\ast}$ satisfies the following equation of Picard-Fuchs type
\begin{equation}
\xi \CZ_\ast'' +\CZ_\ast'-\xi \CZ_\ast = 0\,,
\label{eq:PFDegTor}
\end{equation}
In the analogy to QFT, one can think of Picard-Fuchs equations as analogs of Schwinger-Dyson equations.  In this particular case, the Picard-Fuchs equation happens to be the Bessel equation, and its exact solutions are Bessel functions:
\begin{equation}
\CZ_\ast = C_1 I_0(\xi) + C_2 K_0(\xi)\,.
\end{equation}
Thus we see that \eqref{eq:ParFuncReal1} is one of the fundamental solutions of the Picard-Fuchs equation, and we also explicitly verify that there is precisely one other non-trivial cycle in the $x$-plane.

The above-mentioned surface where the form $\lambda$ lives can be thought of as a double sheeted structure such that each sheet is an infinite strip. Indeed, since real part of $x$ is bounded, $-1<\mathfrak{R}e (x)<1$, we can imagine the surface as two strips $-1<\mathfrak{R}e(x)<1$ connected by a branch cut which connects the points $1$ and $-1$. The A-cycle goes along the cut, whereas the B-cycle goes from plus infinity on the first sheet, passes through the cut and continues back to plus infinity on the second sheet. 
Somewhat less trivially one can represent this surface as a torus\footnote{If instead of putting a cut from $-1$ to $1$ we put two complementary cuts along the real $x$-axis: $(-\infty,-1)$ and $(1,+\infty)$ then we can visualize a torus after completing both $x$-planes with points at infinities. In this, somewhat dual description, integration over the new A-cycle gives $K_0$ and B-cycle integral (which passes through both cuts) yields $I_0$. One should not be worried about the essential singularity in the exponent as it gets `excised' by a branch cut and the contour never approaches it.}. 

Now we are ready to look more carefully at the analytic structure of the $\xi$-plane. There are two apparent singularities: $\xi =0$ and $\xi=\infty$.  Working near the $\xi = 0$ singularity, which corresponds to the strong-coupling limit, one can show that solutions of \eqref{eq:PFDegTor} (which are Bessel functions) behave as 
\begin{align}
K_0(\xi)&= -(f(\xi)+\gamma_{E})\log\left(\frac{\xi}{2}\right) + g(\xi)\,,\notag\\
I_0(\xi)&= f(\xi) \,,
\label{eq:K0I0Res}
\end{align}
where
\begin{equation}
f(\xi) = -1-\frac{\xi^2}{4} + O(\xi^4)\,,
\end{equation}
$g(\xi)$ is another power series in $\xi$, and $\gamma_{E}$ is the Euler-Mascheroni constant. Remarkably, as one can easily check order by order in an expansion of \eqref{eq:PFDegTor} in $\xi$, the same function $f(\xi)$ appears in both expressions. This suggests nice monodromy properties for the solutions. Indeed, because of the logarithm in \eqref{eq:K0I0Res}, the solutions must have the monodromy property 
\begin{equation}
K_0(e^{\pi i}\xi) = f(\xi)\log\xi - \pi i f(\xi) = K_0(\xi)-\pi i I_0(\xi)\,.
\end{equation}
This is of course an implementation of a well-known identity for Bessel functions.  However, we have reproduced this identity directly from the monodromy properties of solutions to \eqref{eq:PFDegTor} around $\xi=0$.  This means that recognizing \eqref{eq:PFDegTor} as a Bessel equation is not important, and our method can be used in more generic situations where the partition function of a theory cannot be written via known special functions, or one does not know the entire strong-coupling expansion. 
Moreover, by choosing the coefficients $C_1$ and $C_2$ appropriately, we can make the monodromy matrix around $\xi=0$ integer-valued. For example, defining 
\begin{equation}
\CZ_{\ast\,A} (\xi)= K_0(\xi)\,,\quad \CZ_{\ast\,B}=\frac{1}{\pi i} I_0(\xi)\,,
\end{equation}
we find that
\begin{equation}
\CZ(\xi,c_1,c_2) = c_1 \CZ_A(\xi) + c_2\CZ_B(\xi)\,,
\label{eq:CZintegerSolution}
\end{equation}
with integer $c_{1,2}$
will give an integral of $\lambda$ over a cycle $\gamma = c_1 A + c_2 B$. We have restored the original partition function $\CZ$ in this formula. Note the extreme similarity of the above formula to \eqref{eq:Ztranseries}. From \eqref{eq:K0I0Res} we see that matrix
\begin{equation}
M_0 = \begin{pmatrix}
1 & -2 \\
0 & 1
\end{pmatrix}
\end{equation}
describes the monodromy properties of period integrals $\CZ_{A,B}$ at $\xi=0$. In other words, if one goes around $\xi=0$ point the A-period integral changes as follows
\begin{equation}
\CZ_A \to \CZ_A - 2\CZ_B\,,
\label{eq:Mondromy00}
\end{equation}
and the $\CZ_B$ integral does not change. 

We emphasize that on the strong-coupling side of the problem, these jumps appear even though all the series are \emph{convergent}.  In fact, we can summarize the situation by saying that at strong coupling our toy model can be naturally represented by transseries built from formal \emph{convergent} series in $\xi$ and logarithms $\log (\xi/2)$, while at weak coupling it is naturally represented in terms of transseries built from formal \emph{divergent} series in $g \sim 1/\xi$ and `instanton' factors $e^{-1/g} \sim e^{-\xi}$. 

The second singularity at $\xi=\infty$ is of essential type, where the solution with $I_0$ blows up exponentially and the other solution goes to zero. It corresponds to the weak coupling regime (small $g$) which we have considered previously. From an analytic perspective, deriving the monodromy properties of the solutions appears to be much easier on the strong coupling side, since it does not require one to deal with resummations. 

Let us now compare monodromy matrix $M_0$ above with $M_{g=0}$ from \eqref{eq:Monodromyg0PerCase}. They certainly look different at first glance. However, if one computes characteristic polynomials for both matrices one sees an agreement up to a sign, which appears due to the opposite orientations of Lefshetz thimbles $\mathcal{J}_{0,1}$ and cycles $A$ and $B$. We can now see that the basis of Lefshetz thimbles $\mathcal{J}_{0,1}$, which is natural for weak-coupling perturbation theory, is in one-to-one correspondence (up to an orientation) with the basis of cycles $A$ and $B$ natural for strong coupling perturbation theory. In particular, the resurgence parameters $\sigma_{0,1}$ \eqref{eq:Ztranseries} are mapped onto the constants $c_{1,2}$ from \eqref{eq:CZintegerSolution}. We may therefore conclude that the monodromy properties of $\CZ(\xi)$ are the same at strong and weak coupling. Thus using the strong coupling analysis we can essentially derive the resurgence properties of the theory at weak coupling.

\section{Elliptic Potential}

In this section we consider a more general exponential integral, with an action which is an elliptic function, which was previously extensively discussed in \cite{Basar:2013eka}.  The integral is
\begin{align}
\mathcal{Z}(g, \zeta) = \frac{1}{g\sqrt{\pi}} \int_{\mathbb{-K}}^{\mathbb{K}}d\phi\, e^{-\frac{1}{g} \text{sd}^2(\phi \vert \zeta) }.
\end{align}
Here $\zeta \in [0,1]$, $\mathbb{K}(\zeta) = \int_0^{\pi/2} dx 1/\sqrt{1-\zeta \sin^2(x)}$ is a complete elliptic integral of the first kind, and $\text{sd}(\phi \vert \zeta)$ is a Jacobi elliptic function.  This is a generalization of our example in the preceding section, since $\text{sd}(\phi \vert 0) = \sin(\phi)$, while $\text{sd}(\phi \vert 1) = \sinh(\phi)$.     For $0<\zeta < 1$ the integral has three saddle-points, two of which coalesce at the degeneration points $\zeta = 0,1$.  

While for e.g. $\zeta = 0$ the integral can be expressed in terms of standard functions via \eqref{eq:SolParFuncReal}, for generic $\zeta$ we do not know of any such closed-form expression for $\mathcal{Z}(g,\zeta)$. The next section briefly reviews the weak-coupling resurgence properties of $\mathcal{Z}(g \vert \zeta)$.  This is a followed by a section giving the  derivation of these properties from a strong-coupling holomorphic analysis.

\subsection{Weak Coupling Approach}
The derivation of the resurgent transseries representation of $\mathcal{Z}(g, \zeta)$ is completely analogous to the one from Section \ref{Sec:TrigWeakCoupling}.  Since it already appeared in \cite{Basar:2013eka} we only briefly sketch it in this section. First, one must find the saddle points of the integral.  Working in an expansion in $g$, one can then compute the formal power series which describe the fluctuations around these saddle points. The large-order behavior of these formal series determines the Stokes automorphisms, which are used to infer the piece-wise constant transseries parameters throughout the complex $g$ plane.  

The saddle points of $\mathcal{Z}(g, \zeta)$ are located at $\phi_i = 0, \mathbb{K}, i\mathbb{K}'$, where $\mathbb{K}'(\zeta) \equiv \mathbb{K}(\zeta')$ with 
$\zeta' \equiv 1-\zeta$.  The actions of these saddle-points are $S_i = 0, 1/\zeta', -1/\zeta$ respectively.  We will refer to these saddle-points as the $i = A, B, C$ saddles, respectively.

To compute the perturbative expansions around the $A$ saddle point, it is convenient to move to a `canonically-normalized' integration variable
\begin{align}
\mathcal{Z}(g, \zeta) = \frac{1}{\sqrt{\pi}} \int_{\mathbb{-K}/\sqrt{g}}^{\mathbb{K}/\sqrt{g}}d\varphi e^{-\frac{1}{g} \text{sd}^2(\varphi \sqrt{g} \vert \zeta) }.
\end{align}
so that the $g^0$ term in an expansion of the integrand is a $g$-independent Gaussian function of $\varphi$.  Expanding in $g$ and integrating term by term on the interval $\varphi \in (-\infty, +\infty)$, one can then show that the fluctuations around the saddle point $A$ take the form
\begin{align}
\label{eq:ASaddleSeries}
\mathcal{Z}_A(g, \zeta) &= 1+  \frac{g}{4} (1-2 \zeta)+ \frac{3g^2 }{32} (8 (\zeta-1) \zeta+3)  \\
&-  \frac{g^3}{128} 15  \left[(2 \zeta-1) (8 (\zeta-1) \zeta+5)\right]+\frac{105 g^4  \left[32 (\zeta-1) \zeta (4 (\zeta-1) \zeta+5)+35\right]}{2048} +\mathcal{O}(g^{5}) \nonumber
\end{align} 

To compute the perturbative expansions around the $B$ and $C$ saddles, we must move to $\varphi$ via $\phi = \phi_{B,C}(\zeta) + \varphi g$.  Expanding in $g$ then gives
\begin{align}
\label{eq:BSaddleSeries}
\mathcal{Z}_B(g, \zeta) &= e^{-S_B} \sqrt{1-\zeta} \left[ 1+\frac{g }{4} \left(1-\zeta^2\right)+\frac{3g^2}{32}  (\zeta-1)^2 (\zeta (3 \zeta+2)+3) \right. \\
&\left.-\frac{15g^3}{128}  (-1 + \zeta)^3 (5 + 3 \zeta + 3 \zeta^2 + 5 \zeta^3)  +\mathcal{O}(g^{4}) \right] \nonumber
\end{align} 
and
\begin{align}
\label{eq:CSaddleSeries}
\mathcal{Z}_C(g, \zeta) &= e^{-S_C} \sqrt{\zeta} \left[1 +\frac{1}{4} g (m-2) m +\frac{3}{32} g^2 m^2 \left(3 m^2-8 m+8\right)\right.\\
&\left. +\frac{15}{128} g^3 m^3 \left(5 m^3-18 m^2+24 m-16\right) +\mathcal{O}(g^{4}) \right] \nonumber
\end{align} 
Note that the C saddle has an exponentially large action when $|g| \ll 1$ and $\arg g \to 0$, which leads to interesting subtleties discussed in \cite{Basar:2013eka}.

The transseries representation of $\mathcal{Z}(g,\zeta)$ is
\begin{align}
\mathcal{Z}(g,\zeta; \sigma_i) = \sum_{i \in \{A,B,C\}} \sigma_i \mathcal{Z}_i(g, \zeta).
\end{align}
It was shown in \cite{Basar:2013eka} that if one sets $\sigma_A =1$, then $\sigma_B = -i, \sigma_C = 0$ for $0 <\arg g <\pi$, and  $\sigma_B = +i, \sigma_C = 0$ for $-\pi < \arg g <0$.  

These Stokes multipliers can be determined as in Section \ref{Sec:TrigWeakCoupling} by computing the jumps in Borel resummation of e.g. 
$\mathcal{Z}_A(g, \zeta)$ as $g$ crosses the Stokes rays $g \in \mathbb{R}^{+}$ and  $g \in \mathbb{R}^{-}$.  These jumps are tied to the large-order behavior of the expansion of $\mathcal{Z}_A(g, \zeta) = \sum_{n=0}^{\infty} a_n g^{n}$, which reads
\begin{align}
a_n \sim \frac{(n-1)!}{\pi} \left[\frac{1}{S_B^{n+1/2}} + \frac{(-1)^n}{(|S_C|)^{n+1/2}}\right], \qquad n \gg 1
\end{align}
Note that the presence of $S_B^{n+1/2}$ and $(|S_C|)^{n+1/2}$  (as opposed to e.g. $S_B^{n}$)  fits nicely with the presence of the overall factors of $\sqrt{\zeta}$ and $\sqrt{\zeta'}$ in our \eqref{eq:BSaddleSeries} and \eqref{eq:CSaddleSeries} above, once one takes into account that the high-order behavior of the series around the $A$ saddle is reflected in the low-order behavior of the expansions around the $B$ and $C$ saddles.  

Given the close match in the structure of the Stokes multipliers to our discussion in Section \ref{Sec:TrigWeakCoupling}, and the fact that the relevant derivations were discussed exhaustively in \cite{Basar:2013eka} (with an emphasis on the interesting role of the complex saddle $C$), we do not repeat our weak-coupling analyses here, and merely quote the relevant result.   Multiplying together the three Stokes matrices found in \cite{Basar:2013eka} (also see \cite{ABSU}), we find that the monodromy matrix associated to the Lefshetz thimble cycles associated to the weak-coupling transseries representation of $\mathcal{Z}(g^2, \zeta)$ is 
\begin{equation}
M_0 = \begin{pmatrix}
1 & 2 & -2 \\
-2 & -3 &2 \\
-2 & -2 & 1
\end{pmatrix}\,.
\label{eq:WeakMonodromy}
\end{equation}

\subsection{Holomorphic Analysis at Strong Coupling}
In the previous example we computed the `quantum moduli space' of the 0d QFT partition function which was parameterized by its coupling constant $\xi$. This moduli space appeared to be a singular torus fibration over the single-ramified complex $\xi$-plane. At $\xi=0$ one of the cycles of the torus vanished. The integral \eqref{eq:ParFuncReal} involved two saddle points.  Our goal here is to discuss the consequences of the modification of \eqref{eq:ParFuncReal} to the following integral
\begin{equation}
\mathcal{Z}(\xi,\zeta) = \int e^{\xi\text{sd}^2(\phi\vert \zeta)} d\phi\,,
\end{equation}
which has three saddle points, rather than two.  Again, the coupling constant $g$ is given by $\xi=-1/2g$.
An integral with a trigonometric function in the exponent is reproduced when $\zeta=0$. After changing the variables as $-x=\text{sd}^2(\phi\vert \zeta)$ we arrive at 
\begin{equation}
\mathcal{Z}(\xi,\zeta) = -\int\limits_{\mathcal{C}} \frac{e^{-\xi x}}{\sqrt{x(1-(1-\zeta)x)(1+\zeta x)}} dx\,.
\end{equation}
Similarly to Section \ref{Sec:StrongTrig}, by differentiating the above integrand with respect to $\xi$ several times, we derive the corresponding Picard-Fuchs equation. We find that
\begin{align}
&\frac{d}{dx}\left(e^{-\xi x}\sqrt{x(1-(1-\zeta)x)(1+\zeta x)}\right) = \notag\\
&\frac{2 (1-\zeta) \zeta  \xi\,z^3  +(2 (1-2 \zeta ) \xi -3 (1-\zeta ) \zeta)\,z^2+(4
   \zeta -2 \xi -2)\,z+1}{\sqrt{x(1-(1-\zeta)x)(1+\zeta x)}}e^{-\xi x}\,,
\end{align}
which leads to the following PF equation of degree three
\begin{equation}
2 (\zeta -1) \zeta  \xi \CZ'''+(2 (1-2 \zeta ) \xi -3 (1-\zeta ) \zeta) \CZ'-(4\zeta -2 \xi -2)\CZ' + \CZ =0\,. 
\label{eq:PFEllPot}
\end{equation}
There are two special values of $\zeta$ when the equation becomes of degree two. When $\zeta=0$, the above equation is reduced to
\begin{equation}
\xi \CZ''+(1+\xi) \CZ' +\frac{1}{2}\CZ =0\,. 
\label{eq:PFEllPotzeta0}
\end{equation}
which is solved by
\begin{equation}
\CZ = e^{-\xi/2}\left(a_1 I_0\left(\frac{\xi}{2}\right)+a_2 K_0\left(\frac{\xi}{2}\right)\right)\,,
\label{eq:PFEllPotSol1}
\end{equation}
Since $\text{sd}^2(x|0)=\sin^2(x)$, here the integral reduces to \eqref{eq:ParFuncReal}. When $\zeta =1$, \eqref{eq:PFEllPot} becomes
\begin{equation}
\xi \CZ''+(1-\xi) \CZ' -\frac{1}{2}\CZ =0\,. 
\label{eq:PFEllPotzeta1}
\end{equation}
whose solutions are 
\begin{equation}
\CZ = e^{\xi/2}\left(b_1 I_0\left(\frac{\xi}{2}\right)+b_2 K_0\left(\frac{\xi}{2}\right)\right)\,.
\end{equation}
Note, however, that apart from these two degenerate cases (when $\zeta\neq0, 1$), the PF equation \eqref{eq:PFEllPot} is of the third order, which suggests that there is another nontrivial cycle which vanishes at the above mentioned values of $\zeta$. We can see from the change of variable we have made that the extra period comes from a pole in the elliptic function. Indeed, $\text{sd}^2(\phi|\zeta)$ has one pole in half of the fundamental domain in the direction of the sum of two quarter periods $\mathbb{K}(\zeta)$ and $\mathbb{K}'(\zeta)$ of this elliptic function. 

Another exactly soluble limit is when $\zeta$ is large, allowing one to drop everything except the leading terms in $\zeta$ in \eqref{eq:PFEllPot}. In this limit two zeros of the denominator in the integrand collide. Interestingly, this regime provides us with a logarithmic cut from the origin in the $1/\xi$ coordinates. The solution behaves as 
\begin{equation}
\CZ\sim e^{\frac{\xi}{\zeta}}\left(C_1+\zeta C_2 + \zeta C_3\log\left(4e^\gamma\frac{\xi}{\zeta}\right)+\dots\right)\,,
\end{equation}
where the ellipses stand for higher order corrections in $1/\xi$. Therefore two solutions scale with $\zeta$ and transform into each other under monodromies around $\xi=\infty$. Thus at weak coupling (recall that $\xi=-1/(2g)$), there is a log-type behavior for one of the period integrals.

At $\zeta=\frac{1}{2}$, equation \eqref{eq:PFEllPot} becomes 
\begin{equation}
-\frac{1}{2} \xi  \CZ^{(3)}(\xi )-\frac{3 \CZ''(\xi )}{4}+2 \xi  \CZ'(\xi )+\CZ(\xi)=0\,,
\end{equation}
and, somewhat surprisingly, it can be solved analytically. The solution turns out to be quadratic in Bessel functions:
\begin{equation}
\CZ= \sqrt{\xi}\left(C_1 I_\frac{1}{4}^2(-\xi)+C_2 I_\frac{1}{4}(-\xi) K_\frac{1}{4}(-\xi)+C_3 K_\frac{1}{4}^2(-\xi)\right)\,.
\label{eq:EllSol14}
\end{equation}
This solution has a square root type branch cut along the positive real axis. In general, when \eqref{eq:PFEllPot} cannot be solved analytically, one should expand it near $\xi=0$ and extract the leading singular behavior of the solutions. This is enough to establish the desired monodromy properties. Indeed, for small $\xi$ one can check that
\begin{align}
K_{\frac{1}{4}}(\xi) &= \xi^{1/4}\frac{\Gamma\left(-\frac{1}{4}\right)}{2^{5/4}}f(\xi)+\xi^{-1/4}\frac{\Gamma\left(\frac{1}{4}\right)}{2^{3/4}}g(\xi)\,,\notag\\
I_{\frac{1}{4}}(\xi ) &= \xi^{1/4}\frac{1}{2^{1/4}\Gamma\left(\frac{5}{4}\right)}f(\xi)\,,
\end{align}
where $f(\xi)$ and $g(\xi)$ are the following series in $\xi$:
\begin{align}
f(\xi) &= 1+\frac{\xi ^2}{5}+\frac{\xi ^4}{90}+\frac{\xi ^6}{3510}+\dots\,,\notag\\
g(\xi) &= 1+\frac{\xi ^2}{3}+\frac{\xi ^4}{42}+\frac{\xi ^6}{1386}+\dots\,.
\end{align}
Note that in this series only even powers of $\xi$ appear.
Hence it is more convenient to define 
\begin{align}
\CZ_1(\xi)=\xi^{1/4}K_{\frac{1}{4}}(\xi) &= \sqrt{\xi}\frac{\Gamma\left(-\frac{1}{4}\right)}{2^{5/4}}f(\xi)+\frac{\Gamma\left(\frac{1}{4}\right)}{2^{3/4}}g(\xi)\,,\notag\\
\CZ_2(\xi)=\sqrt{2}\pi\xi^{1/4}I_{\frac{1}{4}}(\xi ) &= \sqrt{\xi}\frac{2^{1/4}\pi}{\Gamma\left(\frac{5}{4}\right)}f(\xi)\,.
\end{align} 
We can therefore determine the monodromy properties of the above two functions as we go around the branching point $\xi=0$,
\begin{align}
\CZ_1(e^{2\pi i}\xi) &= -\sqrt{\xi}\frac{\Gamma\left(-\frac{1}{4}\right)}{2^{5/4}}f(\xi)+\frac{\Gamma\left(\frac{1}{4}\right)}{2^{3/4}}g(\xi)=S_1(\xi)-2S_2(\xi) \,,\notag\\
\CZ_2(e^{2\pi i}\xi) &= -\sqrt{\xi}\frac{1}{2^{1/4}\Gamma\left(\frac{5}{4}\right)}f(\xi)=-S_2(\xi)\,,
\label{eq:MonodromyMap}
\end{align} 
which leads us to the following monodromy matrix
\begin{equation}
m_0 = \begin{pmatrix}
1 & -2 \\
0 & -1
\end{pmatrix}\,.
\end{equation}
Again the above analytic properties are equivalent to a known reflection formula for Bessel functions
\begin{equation}
K_{\frac{1}{4}}(-\xi)=e^{-\frac{i\pi}{4}} K_{\frac{1}{4}}(\xi)-i \pi  I_{\frac{1}{4}}(\xi )\,.
\end{equation}
To understand the monodromy properties of the full solution, it is useful to observe that a vector in the space of solutions \eqref{eq:EllSol14} can be viewed as an image of the Veronese map from $\mathbb{C}^2$ generated by $S_1$ and $S_2$ to $\mathbb{C}^3$.  Indeed, applying \eqref{eq:MonodromyMap} to \eqref{eq:EllSol14} we get 
\begin{equation}
M_0 = \begin{pmatrix}
1 & 4 & 4 \\
0 & -1 &-2 \\
0 & 0 & 1
\end{pmatrix}\,.
\label{eq:EllMonodMatrix}
\end{equation}
Again, we can see that the characteristic polynomial of the above matrix coincides with the one for the weak monodromy matrix \eqref{eq:WeakMonodromy} (up to a change of orientation). 

More generally, when $\zeta\neq 1/2$, one can still derive the formal monodromy properties of \eqref{eq:PFEllPot}. Indeed, we can rewrite the third order ODE \eqref{eq:PFEllPot} as a first order matrix equation with respect to the vector $\vec{\CZ}=(\CZ\, \CZ'\, \CZ'')^T$
\begin{equation}
\vec{\CZ}' = A \vec{\CZ}\,, \qquad
A = \left(
\begin{array}{ccc}
 0 & 1 & 0 \\
 0 & 0 & 1 \\
 -\frac{1}{2 (\zeta -1) \zeta  \xi } & -\frac{-2 \zeta +\xi +1}{(\zeta -1) \zeta  \xi
   } & -\frac{1-2 \zeta }{(\zeta -1) \zeta }-\frac{3}{2 \xi } \\
\end{array}
\right).
\end{equation}
Note that at small $\xi$, the matrix $A$ has a simple pole (regular singular point) $A\sim A_0/\xi$, so we should expect a power-law-type branching at $\xi=0$\footnote{Also log-branching is possible for regular singular points, but it is not realized in this example for generic $\zeta$. It happens though that when $\zeta=0$ we have a log-type branching.}. Formally the solution can be written as a matrix exponential. In order to derive the monodromy matrix we need to look at the behavior of the eigenvalues of matrix $A$ near $\xi=0$. We find that the only nonzero eigenvalue is independent of  $\zeta$ and equal to $-3/2$, meaning that \eqref{eq:EllMonodMatrix} describes the monodromy properties of the integration cycles for any value of $\zeta$. 

Therefore, just as in the previous section, we observe a direct match between the monodromy properties at weak and strong coupling.

\section{Conclusions and Future Directions}
In this paper we discussed  the use of resurgence theory to connect weak-coupling and strong-coupling results, and began an exploration of a potentially promising correspondence between resurgence and holomorphy.   We were able to make three main observations in the 0d toy models that we have considered.

First, we emphasized that naive weak-coupling perturbative expansions can fail even when $|g| \ll 1$ if $\arg g \neq 0$.  This happens despite the fact that the perturbative power series satisfies the standard criterion for its sensibility of having its subsequent terms grow smaller as $|g|$ is decreased.  We showed that the cure for this problem is taking into account the contribution of \emph{all} of the saddle points in the problem via the technology of resurgent transseries.   

Of course, naive weak-coupling perturbation theory also fails to accurately describe what happens for $|g| \gtrsim 1$.  Our second result is that the cure for this issue is in fact the same as the cure for the problems at $|g| \ll 1$.  Once the information about all of the saddle points is taken into account using resurgence theory, one can make highly accurate predictions for the $|g| \gtrsim 1$ behavior using \emph{only} weak-coupling data.     

Our third observation is that the different terms in resurgent transseries \eqref{eq:Ztranseries}, which formally represent `quantum' fluctuations around perturbative and non-perturbative saddle points of our toy-model integrals, can be identified with a basis of integration cycles on a singular surface parameterized by the complexified coupling constant. While the natural home of resurgence theory is at weak coupling,  the holomorphic Picard-Fuchs approach appears to be more natural at strong coupling. By working out the holomorphic dependence of the partition function on the coupling constant at strong coupling, we were able to derive a differential equation of Picard-Fuchs type which is solved by the partition function. From this equation we extracted the monodromy properties of the period integrals, which are mapped onto the resurgence relations between the perturbative and the non-perturbative series expansions of the partition function at weak coupling. 

Our weak-coupling and strong-coupling analyses should be extended to multi-dimensional matrix integrals which appear in matrix models \cite{Marino:2012zq} and supersymmetric localization computations. Indeed, theories with extended supersymmetry are known to enjoy localization properties: their path integrals localize on certain field configurations, which reduce the path integrals to finite dimensional matrix-type integrals. So one may study e.g. the Picard-Fuchs equations for these partition functions for supersymmetric theories in various dimensions in supersymmetry preserving backgrounds (Omega-background, spheres, etc.). 

The next natural step is to consider the spectra of quantum mechanical systems where non-perturbative effects play a crucial role.   In QM (and also in QFT) there are typically an \emph{infinite} number of distinct (quasi)-saddle-points.  As a result, the generalization of our weak-coupling analysis to QM is likely to be non-trivial.  It will also be interesting to work out the relation between resurgence theory and holomorphy techniques in QM and QFT.  Some work has already been done in this direction \cite{Gulden:2013wya, Gorsky:2014lia, Krefl:2013bsa}.   For example, if we study a particle in potential with harmonic-type degeneracy (double-well potential, sine-Gordon potential, etc) quantum mechanical instantons contribute to non-perturbative corrections to the spectrum by splitting the degeneracies or forming Bloch bands. In order to obtain an exact spectrum \cite{ZinnJustin:2004cg,Jentschura:2010zza,ZinnJustin:2010ng,Jentschura:2011zza,Dunne:2014bca} one needs to sum all instanton corrections which is undoubtedly a formidable task. An exact formula (exact WKB quantization), which was conjectured in the papers just cited and proven in \cite{Dunne:2014bca}, exists, but an understanding of the resurgence mechanism in general from the strong-coupling holomorphy perspective is not yet available. Our hope is that the techniques outlined in this paper will help understand these issues more deeply.

\section*{Acknowledgements}
We thank Michael Janas, who collaborated with us on the initial stage of the project. In addition, we are grateful to G\"{o}k\c{c}e Ba\c{s}ar, Gerald Dunne, and Arkady Vainshtein for fruitful discussions. PK would like to thank W. Fine Theoretical Physics Institute at University of Minnesota, University of California, Berkeley, as well as Simons Center for Geometry and Physics at Stony Brook University for kind hospitality, where part of his work was done. The research of PK was supported by the Perimeter Institute for Theoretical Physics. Research at Perimeter Institute is supported by the Government of Canada through Industry Canada and by the Province of Ontario through the Ministry of Economic Development and Innovation. AC thanks the US Department of Energy for support under grant number DE-SC0011842.

\bibliography{res1}

\end{document}